\documentclass[twocolumn,showpacs,prb]{revtex4}

\usepackage{graphicx}    
\usepackage{dcolumn}     
\usepackage{bm}          
\usepackage{amsmath}

\begin{document}
\preprint{}

\title{Theory of proximity effect in  
normal metal/$d_{x^2-y^2}$-wave superconductor interface 
in the presence of subdominant components of the pair potentials
}

\author{
Y. Tanuma,$^1$
Y. Tanaka,$^{2,3}$ and S. Kashiwaya$^4$
}%
%
\affiliation{
$^1$Institute of Physics, Kanagawa University,
Rokkakubashi, Yokohama, 221-8686, Japan \\
$^2$Department of Applied Physics, 
Nagoya University, Nagoya, 464-8603, Japan \\
$^3$CREST, Japan Science and Technology Cooperation (JST),
Nagoya, 464-8603, Japan  \\
$^4$National Institute of Advanced Industrial Science
and Technology, Tsukuba, 305-8568, Japan
}
%
\date{\today}

\begin{abstract}
Superconducting proximity effect in
normal metal (N) / $d_{x^2-y^2}$-wave superconductor (D)
junctions in the presence of attractive interelectron potentials
which can induce subdominant $s$-wave pair potentials
both in N and D sides,
is studied based on the quasiclassical 
Green's function theory,
where spatial dependencies of the pair potentials are 
determined self-consistently. 
In the N/D junctions with orientational angle with $\theta = 0$, 
the $s$-wave component is induced in the N side
by the proximity effect only for high transparent case,
where the induced $s$-wave components in both the N and D sides
do not break the time reversal symmetry (TRS).
For fully transparent case, 
the resulting local density of states 
has a very sharp zero-energy peak (ZEP),
the origin of which is the sign change of the pair potentials
felt by the quasiparticles between the $s$-wave component in the N side
and $d_{x^2-y^2}$-wave dominant component in the D side
through Andreev reflections. 
On the other hands, for $\theta = \pi/4$,
the subdominant $s$-wave component which breaks the TRS 
appears near the interface.
Besides, for lower transparent cases,
the subdominant imaginary $s$-wave component 
is also induced near the interface in the N side.
The proximity induced $s$-wave component in the N side
does not enhance the magnitude of the $s$-wave component of 
the pair potential which break the TRS in the D side. 
The resulting LDOS at the interface has the ZEP or its splitting
depending on the transparency of the junction.
\end{abstract}

\maketitle

\section{Introduction}
To determine the pairing symmetry in unconventional
superconductors is an interesting problem to
understand the pairing mechanism of superconductivity.
Nowadays, it is widely accepted that
the superconducting pair potential of high-$T_c$ cuprates
has a $d_{x^{2}-y^{2}}$-wave symmetry
in the bulk state.
\cite{SR92,SR95,Scalapino,Harlin,Tsuei}
In order to determine the pairing symmetry,
several phase-sensitive probes have been used
\cite{SR92,SR95,Harlin,Tsuei}.
Among them tunneling spectroscopy via
Andreev bound states (ABS's),
\cite{Buch,Hu,TK95,KT95,Matsu1,FRS97,KT00,KT96,Barash,Lofwander,ATS04}
help us to detect the internal phase
in the pair potential.
The formation of ABS's at the Fermi energy (zero-energy),
which is originated from the interference effect
in the effective pair potential of
$d_{x^{2}-y^{2}}$-wave symmetry
through the reflection at the surface/interface,
plays an important role 
when the angle $\theta$ between the lobe direction of the
$d_{x^{2}-y^{2}}$-wave pair potential
and the normal to the surface/interface is nonzero.
\cite{Hu}
In fact, the existence of ABS's,
which manifests itself as a distinct conductance peak
at zero-bias in the tunneling spectrum (ZBCP),
has been actually observed.
Up to now, the consistency between theories
\cite{Tanu1,Tanu2,Tanu2',Zhu,W1,W2,asa1,asa2,r1,r2,r3,r4,r5,r6,r7,r8,r9,Lubi03}
and experiments
\cite{e1,e2,e3,e4,e5,e6,e7,e8,e9,e10,Koren,Boston,e11,e12,e13}
has been checked in detail.
\par
On the other hand, 
at the surface/interface of $d_{x^{2}-y^{2}}$-wave 
superconductor, it is known 
that the time-reversal symmetry (TRS) may be broken.
\cite{SBL95,Kuboki,Sig98}
The reduction of the amplitude 
of $d_{x^{2}-y^{2}}$-wave pair potential near the surface/interface
at low temperatures 
allows to induce the subdominant
$s$-wave [$d_{xy}$-wave] component
in the imaginary part of the pair potential,
i.e.,
$d$+i$s$-wave \cite{FRS97,Matsu2,Matsu3,Tanu2,Tanu2',Zhu}
[$d$+i$d'$-wave \cite{Krishana,Balatsky,Laughlin} ] state.
The amplitude of the induced $s$-wave component 
has a maximum at $\theta =\pi/4$, which 
blocks the motion of quasiparticles. 
Then, the energy levels of ABS's shift from zero
and the resulting tunneling spectra has a ZBCP splitting
even in zero magnetic field.
The observation of the ZBCP splitting without magnetic fields
was believed to be one of the evidence
for the broken time reversal symmetry states (BTRSS).
However, actual tunneling experiments still
remain to be controversial.
In fact, some groups \cite{e5,e6,e9,e10,Koren,Boston}
have reported the ZBCP splitting
and they ascribed
the origin of the ZBCP splitting
to the above BTRSS. 
At the same time,
there are other experiments which do not show
the ZBCP splitting, and in these experiments,
the ZBCP survives even at low temperature.
\cite{e3,e4,e7,e8,e11,e12,e13}
Moreover, 
a critical current measurement of
grain boundary junctions in high-$T_c$ cuprates
concluded the absence of BTRSS at the interface.
\cite{Neils}
It has been studied in detail that 
the ZBCP splitting  in the tunneling spectra
is sensitive to several factors:
(i) transmission probability of the junctions,
\cite{Tanu5}
(ii) roughness at the interface,
\cite{Tanu1,Tanu2,Tanu2'}
and
(iii) effect of impurity scattering
in normal metals or superconductors.
\cite{r1,asa1}
Recent studies by Asano \textit{et al.},
\cite{asa1} revealed that the existence of the 
impurity scattering can induce the 
ZBCP splitting even without BTRSS.
Taking account of these situations,
one can not conclude that 
the observation of the ZBCP splitting
by tunneling experiments 
directly indicate the presence of the BTRSS
near the surface/interface.
\par
More recently, 
it is proposed that
the induced $s$-wave component of the pair potential 
by proximity effect in normal metals (N) may
enhance the magnitude of 
the subdominant $s$-wave component in
$d_{x^2-y^2}$-wave superconductors (D),
which forms the BTRSS based on the analysis of
the tunneling experiments.
\cite{Kohen}
%
%
It is of great interest to study
whether the induced $s$-wave component
in the N side by the proximity effect has the influence on
the subdominant component in the D side
which forms the BTRSS.
The proximity effect in the N/D junction without the BTRSS
was theoretically studied by Ohashi.
\cite{Ohashi}
It is shown that the amplitude of the subdominant i$s$-wave is
the largest at $\theta = \pi/4$.
For $\theta=0$ with fully transparent junctions,
local density of states (LDOS) at the interface
has a zero energy peak (ZEP), the origin of which is not 
ABS but the sign change of the pair potentials 
between N and D sides felt by
Andreev reflection of quasiparticles.
%
%
It is necessary to study the interplay between the 
proximity effect and the BTRSS by changing the orientational angle $\theta$.
At the same time,
it is also interesting to clarify the relationship
between the different origin of two kinds of
ZEP's.
One is the ZEP which originates from the formation of ABS
for $\theta \neq 0$.
The other is due to the proximity effect expected for
fully transparent junctions for $\theta=0$.
\cite{Ohashi}
\par
In the present paper, 
we study the proximity effect
in N/D junctions on the basis of
quasiclassical Green's function methods.
We assume attractive interelectron potentials 
which induce subdominant $s$-wave components
in the N side as well as the D.
The spatial dependencies of the pair potentials 
both in the N and D sides are determined self-consistently. 
The LDOS at the interface of the N/D junctions 
are studied in detail by changing $\theta$ and 
the transparency of the junction. 
For $\theta = \pi/4$, the magnitude of the 
$s$-wave component is induced even in the N side,
while the subdominant $s$-wave component
which breaks TRS exists near the interface
on the D side.
\cite{Lof}
On the other hand, for $\theta = 0$, 
$s$-wave component is induced in the N side
by the proximity effect with the increase of
the transparency of the junction. 
The induced $s$-wave component in the N side and that
in the D side do not break the TRS. 
It is revealed that the 
proximity induced $s$-wave component in the N side
does not enhance the BTRSS in the D side. 
\par
The organization of the present paper is as follows.
In Sec.~\ref{sec:02},
a theoretical formulation to calculate
the spatial dependence of the pair potential,
and local density of states is presented.
In Sec.~\ref{sec:03},
results of the numerical calculations 
are discussed in detail.
Finally,
we summarize the paper in Sec.~\ref{sec:04}.
\par
%
\begin{figure}[htb]
\begin{center}
\scalebox{0.60}{
\includegraphics{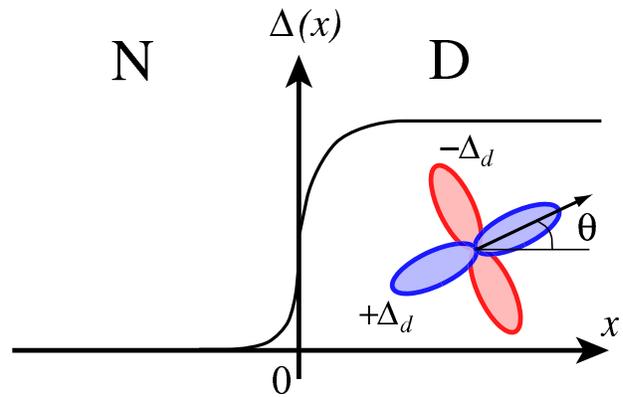}}
\caption{Schematic illustration of a N/D junction
with a spatially varying pair potential. 
The misorientation angle $\theta$ denotes the 
angle between the crystal axis of $d_{x^2-y^2}$-wave superconductor 
and the normal to the interface. 
\label{fig:01}}
\end{center}
\end{figure}
%
\section{Theoretical model}
\label{sec:02}
%
In the present paper, 
we consider the N/D junction
separated by an insulating interface at $x=0$,
where the normal metal is located at $x<0$
and the $d_{x^{2}-y^{2}}$-wave superconductor
extends elsewhere (Fig.~\ref{fig:01}). 
For the simplicity,
two dimensional system is assumed and
the $x$ axis is taken perpendicular to the interface.
When quasiparticles are in the $xy$ plane,
a transmitted electron like quasiparticle
and holelike quasiparticle in the D side feel 
different effective pair potentials
$\Delta_{d}(\phi_{+})$ and $\Delta_{d}(\phi_{-})$,
with $\phi_{+}=\phi$ and $\phi_{-}=\pi-\phi$.
Here $\phi$ is the azimuthal angle
in the $xy$ plane 
given by $(k_{x}+{\rm i}k_{y})/|\bm{k}|={\rm e^{i\phi}}$.
Besides, a cylindrical Fermi surface is assumed 
and the magnitude of the Fermi momentum 
and the effective mass are chosen to be equal 
both in the N and D sides.
\par
In order to study the proximity effect
in the N/D junction,
we determine the spatial variation
of the pair potentials self-consistently.
For this purpose,
we make use of the quasi-classical Green's function procedure
\cite{Eilen} developed by Nagai and co-workers.
\cite{Nagai,Ashida}
In the following,
we briefly summarize this scheme we employed.
We introduce the generalized Eilenberger equation,
\cite{Eilen}
\begin{align}
{\rm i}|v_{{\rm F}x}|\frac{\partial}{\partial x}
g_{\alpha \beta}(\phi,x) = &-\alpha
\left [ {\rm i}\omega_{m}\hat{\tau}_{3}
+\hat{\Delta}(\phi_{\alpha},x) \right ] 
g_{\alpha \beta}(\phi,x)
\nonumber \\
&+ \beta  g_{\alpha \beta}(\phi,x)
\left [ {\rm i}\omega_{m}\hat{\tau}_{3}
+\hat{\Delta}(\phi_{\beta},x) \right ] , \\
\hat{\Delta}(\phi_{\alpha},x) =&
\left (
    \begin{array}{cc}
        0 & \Delta(\phi_{\alpha},x) \\
  -\Delta(\phi_{\alpha},x)^{*} & 0
    \end{array}
\right ),
\end{align}
where $v_{{\rm F}x}=v_{\rm F}\cos \theta$
and $\hat{\tau}_i (i=1,2,3)$ stand for
the $x$ component of the Fermi velocity
and the Pauli matrices, respectively.
Here $\omega_m=\pi T(2m+1)$ ($m$: integer)
is the Matsubara frequency.
\par
%
Now,
considering a semi-infinite N/D junction geometry,
the pair potential in D [N] side
tend to the bulk value [zero]
$\Delta^{\rm D}(\phi_{\alpha},\infty)$
[$\Delta^{\rm N}(\phi_{\alpha},-\infty)$]
at sufficiently large $x$.
In semi-infinite limit,
we can find the quasi-classical Green's function
$\hat{g}_{\alpha \alpha}^{l}(\phi_{\alpha},x)$
for $l(=\textrm{N,D})$ regions given by
\cite{Ohashi,Nagai,Nagato}
\begin{align}
\hat{g}_{\alpha \alpha}^{l}(\phi_{\alpha},x)
={\rm i}
\left (
\frac{2\hat{A}^{l}_{\alpha}(x)}
{{\rm Tr}[\hat{A}^{l}_{\alpha}(x)]}
-\bm{1}
\right ),
\end{align}
Following equations are satisfied 
for the N side,
\begin{align}
\label{an+}
\hat{A}^{\rm N}_{+}(x)
&=
\tilde{U}^{\rm N}_{+}(\phi_{+},x,0)
\hat{R}_{\rm D}
\hat{\lambda}^{\rm N}(0,x),
\\
\hat{A}^{\rm N}_{-}(x)
&=
\hat{\lambda}^{\rm N}(x,0)
\hat{R}_{\rm D}
\tilde{U}^{\rm N}_{-}(\phi_{-},0,x),
\\ \nonumber
\hat{\lambda}^{\rm N}(x,x^{\prime}) &=
\left (
\begin{array}{cc}
                1 &
\mathrm{i}/{\cal G}^{\rm N}_{+}(x^{\prime}) \\
      \mathrm{i}{\cal G}^{\rm N}_{-}(x) &
      -{\cal G}^{\rm N}_{-}(x)/
      {\cal G}^{\rm N}_{+}(x^{\prime}) \\
\end{array}
\right ),
\end{align}
and for the D side,
\begin{align}
\label{ad+}
\hat{A}^{\rm D}_{+}(x)
&=
\hat{\lambda}^{\rm D}(x,0)
\hat{R}_{\rm N}
\tilde{U}^{\rm D}_{+}(\phi_{+},0,x),
\\
\label{ad-}
\hat{A}^{\rm D}_{-}(x)
&=
\tilde{U}^{\rm D}_{-}(\phi_{-},x,0)
\hat{R}_{\rm N}
\hat{\lambda}^{\rm D}(0,x),
\\ \nonumber
\hat{\lambda}^{\rm D}(x,x^{\prime}) &=
\left (
\begin{array}{cc}
                1 &
\mathrm{i}/{\cal G}^{\rm D}_{-}(x^{\prime}) \\
      \mathrm{i}{\cal G}^{\rm D}_{+}(x) &
      -{\cal G}^{\rm D}_{+}(x)/
      {\cal G}^{\rm D}_{-}(x^{\prime}) \\
\end{array}
\right ), 
\end{align}
respectively. 
In the above,
$\hat{R}_{\rm N}$ and $\hat{R}_{\rm D}$ are matrices
with reflection probability $R$ at the interface
given by
\cite{Ohashi}
\begin{align}
\hat{R}_{\rm N} & \propto
\left (
\begin{array}{cc}
  {\cal G}^{\rm N}_{+}(0) - R{\cal G}^{\rm N}_{-}(0) &
            \mathrm{i}(1-R) \\
   \mathrm{i}(1-R){\cal G}^{\rm N}_{-}(0)
                  {\cal G}^{\rm N}_{+}(0) &
  {\cal G}^{\rm N}_{+}(0)R - {\cal G}^{\rm N}_{-}(0) \\
\end{array}
\right ),
\\
\hat{R}_{\rm D} & \propto
\left (
\begin{array}{cc}
  {\cal G}^{\rm D}_{-}(0) - R{\cal G}^{\rm D}_{+}(0) &
            \mathrm{i}(1-R) \\
   \mathrm{i}(1-R){\cal G}^{\rm D}_{+}(0)
                  {\cal G}^{\rm D}_{-}(0) &
  {\cal G}^{\rm D}_{-}(0)R - {\cal G}^{\rm D}_{+}(0) \\
\end{array}
\right ).
\end{align}
Here, we define ${\cal G}^{l}_{\alpha}(x)$ given by
\begin{align}
\label{gg}
{\cal G}^{\rm N}_{\alpha}(x)=
-{\rm i}\frac{v_{n}^{{\rm N}(+)}(\phi_{\alpha},x)}
{u_{n}^{{\rm N}(+)}(\phi_{\alpha},x)}, \quad
{\cal G}^{\rm D}_{\alpha}(x)=
-{\rm i}\frac{v_{n}^{{\rm D}(-)}(\phi_{\alpha},x)}
{u_{n}^{{\rm D}(-)}(\phi_{\alpha},x)}.
\end{align}
\par
Next, in order to obtain the quantities
$\tilde{U}^{l}_{\alpha}(\phi_{\alpha},x,x^{\prime})$
in Eqs.~(\ref{an+})-(\ref{ad-}),
we rewrite $\hat{A}^{l}_{\alpha}(x)$ as
\cite{Nagai,Nagato}
\begin{align}
\hat{A}^{\rm N}_{+}(x) &=
\left (
\begin{array}{c}
      X^{\rm N}_{+}(x) \\
      Y^{\rm N}_{+}(x) \\
\end{array}
\right )
\left (
\begin{array}{cc}
      u_{n}^{{\rm N}(+)}(\phi_{+},x) &
      v_{n}^{{\rm N}(+)}(\phi_{+},x) \\
\end{array}
\right )
\hat{\tau}_{2},
\\
\hat{A}^{\rm N}_{-}(x) &=
\left (
\begin{array}{c}
      u_{n}^{{\rm N}(+)}(\phi_{-},x) \\
      v_{n}^{{\rm N}(+)}(\phi_{-},x) \\
\end{array}
\right )
\left (
\begin{array}{cc}
      X^{\rm N}_{-}(x) &
      Y^{\rm N}_{-}(x) \\
\end{array}
\right )
\hat{\tau}_{2},
\\
\hat{A}^{\rm D}_{+}(x) &=
\left (
\begin{array}{c}
      u_{n}^{{\rm D}(-)}(\phi_{+},x) \\
      v_{n}^{{\rm D}(-)}(\phi_{+},x) \\
\end{array}
\right )
\left (
\begin{array}{cc}
      X^{\rm D}_{+}(x) &
      Y^{\rm D}_{+}(x) \\
\end{array}
\right )
\hat{\tau_{2}},
\\
\hat{A}^{\rm D}_{-}(x) &=
\left (
\begin{array}{c}
      X^{\rm D}_{-}(x) \\
      Y^{\rm D}_{-}(x) \\
\end{array}
\right )
\left (
\begin{array}{cc}
      u_{n}^{{\rm D}(-)}(\phi_{-},x) &
      v_{n}^{{\rm D}(-)}(\phi_{-},x) \\
\end{array}
\right )
\hat{\tau}_{2},
\end{align}
with
\begin{align}
\left (
\begin{array}{c}
      X^{\rm N}_{-}(x) \\
      Y^{\rm N}_{-}(x) \\
\end{array}
\right )
= &
\tilde{U}^{\rm N}_{+}(\phi_{+},x,0)
\hat{R}_{\rm D}
\left (
\begin{array}{c}
      u_{n}^{{\rm N}(+)}(\phi_{-},0) \\
      v_{n}^{{\rm N}(+)}(\phi_{-},0) \\
\end{array}
\right ),
\\
\left (
\begin{array}{cc}
      X^{\rm N}_{-}(x) &
      Y^{\rm N}_{-}(x) \\
\end{array}
\right )
\hat{\tau}_{2} =&
\left (
\begin{array}{cc}
      u_{n}^{{\rm N}(+)}(\phi_{+},0) &
      v_{n}^{{\rm N}(+)}(\phi_{+},0) \\
\end{array}
\right ) \hat{\tau}_{2} \nonumber \\ &\times
\hat{R}_{\rm D}
\tilde{U}^{\rm N}_{-}(\phi_{-},0,x),
\\
\left (
\begin{array}{cc}
      X^{\rm D}_{+}(x) &
      Y^{\rm D}_{+}(x) \\
\end{array}
\right )
\hat{\tau}_{2} =&
\left (
\begin{array}{cc}
      u_{n}^{{\rm D}(-)}(\phi_{-},0) &
      v_{n}^{{\rm D}(-)}(\phi_{-},0) \\
\end{array}
\right ) \hat{\tau}_{2} \nonumber \\ &\times
\hat{R}_{\rm N}
\tilde{U}^{\rm D}_{+}(\phi_{+},0,x),
\\
\left (
\begin{array}{c}
      X^{\rm D}_{-}(x) \\
      Y^{\rm D}_{-}(x) \\
\end{array}
\right )
= &
\tilde{U}^{\rm D}_{-}(\phi_{-},x,0)
\hat{R}_{\rm N}
\left (
\begin{array}{c}
      u_{n}^{{\rm D}(-)}(\phi_{+},0) \\
      v_{n}^{{\rm D}(-)}(\phi_{+},0) \\
\end{array}
\right ).
\end{align}
Then we define ${\cal F}^{l}_{\alpha}(x)$ given by
\begin{align}
\label{ff}
{\cal F}^{\rm N}_{\alpha}(x)=
{\rm i}\frac{X^{\rm N}_{\alpha}(x)}
{Y^{\rm N}_{\alpha}(x)}, \quad
{\cal F}^{\rm D}_{\alpha}(x)=
{\rm i}\frac{X^{\rm D}_{\alpha}(x)}
{Y^{\rm D}_{\alpha}(x)}.
\end{align}
\par

\begin{figure}[htb]
\begin{center}
\scalebox{0.55}{
\includegraphics{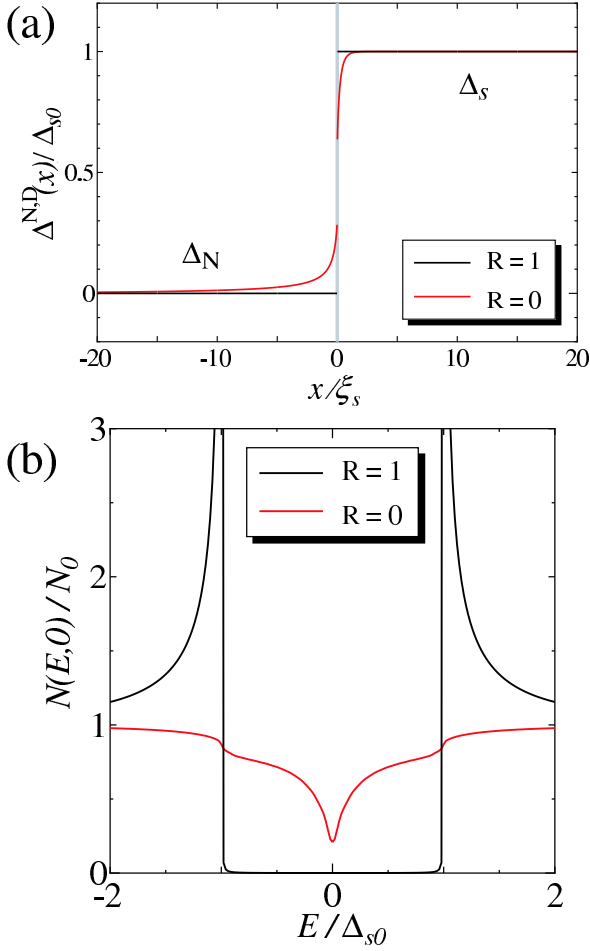}}
\caption{(a) Spatial dependence of the pair potentials 
in a normal metal /$s$-wave superconductor (N/S) junction
with $T=0.02T_{s}$ and $T_{\rm N}/T_{s}=10^{-3}$.
(b) The corresponding local density of states
(LDOS) at the S side of the interface.  
\label{fig:02}}
\end{center}
\end{figure}

%
Here, ${\cal G}^{l}_{\alpha}(x)$ and
${\cal F}^{l}_{\alpha}(x)$
in Eqs.~(\ref{gg}) and~(\ref{ff}), 
obey the following Riccati type equations:
\begin{align}
\label{req}
\hbar |v_{{\rm F}x}|
\frac{\partial}{\partial x}{\cal G}^{l}_{\alpha}(x)
= & \alpha \left [
2\omega_{m} {\cal G}^{l}_{\alpha}(x) +
\Delta^{l}(\phi_{\alpha},x)
{\cal G}^{l}_{\alpha}(x)^{2}
\right .
\nonumber \\
&- \left . \Delta^{l}(\phi_{\alpha},x)^{*}
\right ],
\\
\hbar |v_{{\rm F}x}|
\frac{\partial}{\partial x}{\cal F}^{l}_{\alpha}(x)
= &-\alpha \left [
2\omega_{m} {\cal F}^{l}_{\alpha}(x)
- \Delta^{l}(\phi_{\alpha},x)^{*}
{\cal F}^{l}_{\alpha}(x)^{2}
\right . \nonumber \\
&+ \left . \Delta^{l}(\phi_{\alpha},x)
\right ].
\end{align}
We can write the quasi-classical Green's function
in a compact form \cite{Nagai}
\begin{widetext}
\begin{align}
\label{gn--}
\hat{g}^{\rm N}_{--}(\phi_{-},x) &=
\frac{{\rm i}}{1-{\cal G}^{\rm N}_{-}(x)
{\cal F}^{\rm N}_{-}(x)}
\left (
\begin{array}{cc}
1+{\cal G}^{\rm N}_{-}(x){\cal F}^{\rm N}_{-}(x) &
2{\rm i}{\cal F}^{\rm N}_{-}(x) \\
2{\rm i}{\cal G}^{\rm N}_{-}(x) &
-1-{\cal G}^{\rm N}_{-}(x){\cal F}^{\rm N}_{-}(x)
\\
\end{array}
\right ),
\\
\label{gd++}
\hat{g}^{\rm D}_{++}(\phi_{+},x) &=
\frac{{\rm i}}{{\cal G}^{\rm D}_{+}(x)
{\cal F}^{\rm D}_{+}(x)-1}
\left (
\begin{array}{cc}
{\cal G}^{\rm D}_{+}(x){\cal F}^{\rm D}_{+}(x)+1 &
2{\rm i}{\cal F}^{\rm D}_{+}(x) \\
2{\rm i}{\cal G}^{\rm D}_{+}(x) &
-{\cal G}^{\rm D}_{+}(x){\cal F}^{\rm D}_{+}(x)-1
\\
\end{array}
\right ),
\end{align}
\end{widetext}
with $\hat{g}^{l}_{--}(\phi_{-},x)
=-\hat{g}^{l}_{++}(-\phi_{+},x)^{\dagger}$.
Initial conditions of these equations are as follows, 
\begin{align}
{\cal G}_{-}^{\rm N}(-\infty)=0, \quad
{\cal G}_{\alpha}^{\rm D}(\infty)=
\frac{\Delta^{\rm D}(\phi_{\alpha},\infty)^{*}}
{\omega_{m} + \alpha \Omega_{\alpha}^{\rm D}},
\end{align}
with $\Omega_{\alpha}^{l}=\sqrt{\omega_{m}^{2}
+|\Delta^{l}(\phi_{\alpha},\infty)|^{2}}$.
Moreover, the boundary condition of 
the ${\cal G}_{\alpha}^{l}(0)$ and
${\cal F}_{\alpha}^{l}(0)$
at the interface $x=0$
are 
\begin{align}
\label{fn+1}
{\cal F}^{\rm N}_{+} &=
\frac{{\cal G}^{\rm D}_{-} - R{\cal G}^{\rm D}_{+}
- (1-R){\cal G}^{\rm N}_{-}}
{[R{\cal G}^{\rm D}_{-} - {\cal G}^{\rm D}_{+}]
{\cal G}^{\rm N}_{-}
+(1-R){\cal G}^{\rm D}_{+}{\cal G}^{\rm D}_{-}},
\\
\label{fn-1}
{\cal F}^{\rm N}_{-} &=
\frac{R{\cal G}^{\rm D}_{-} - {\cal G}^{\rm D}_{+}
+ (1-R){\cal G}^{\rm N}_{+}}
{{\cal G}^{\rm N}_{+}
[{\cal G}^{\rm D}_{-} - R{\cal G}^{\rm D}_{+}]
-(1-R){\cal G}^{\rm D}_{-}{\cal G}^{\rm D}_{+}},
\\
\label{fd+2}
{\cal F}^{\rm D}_{+} &=
\frac{R{\cal G}^{\rm N}_{+} - {\cal G}^{\rm N}_{-}
+ (1-R){\cal G}^{\rm D}_{-}}
{{\cal G}^{\rm D}_{-}
[{\cal G}^{\rm N}_{+} - R{\cal G}^{\rm N}_{-}]
-(1-R){\cal G}^{\rm N}_{+}{\cal G}^{\rm N}_{-}},
\\
\label{fn-2}
{\cal F}^{\rm D}_{-} &=
\frac{{\cal G}^{\rm N}_{+} - R{\cal G}^{\rm N}_{-}
- (1-R){\cal G}^{\rm D}_{+}}
{[R{\cal G}^{\rm N}_{+} - {\cal G}^{\rm N}_{-}]
{\cal G}^{\rm D}_{+}
+(1-R){\cal G}^{\rm N}_{-}{\cal G}^{\rm N}_{+}}.
\end{align}
\par
The pair potentials for both N and D sides are given by
\cite{Matsu2,Ohashi,Ashida,Nagato,Bruder}
\begin{align}
\label{peq}
\Delta^{l}(\phi,x) =&
\sum_{0 \leq m < \omega_{c}/2\pi T}
\frac{1}{2\pi}
{\displaystyle
\int ^{\pi/2}_{-\pi/2}
d\phi^{\prime}\sum_{\alpha}
V^{l}(\phi,\phi^{\prime}_{\alpha})}
\nonumber \\
& \times
[\hat{g}^{l}_{\alpha \alpha}(\phi^{\prime}_{\alpha},x)]_{12},
\end{align}
where $\omega_{c}$ is the cutoff energy and
$[\hat{g}^{l}_{\alpha \alpha}(\phi_{\alpha},x)]_{12}$
means the 12 element of
$\hat{g}^{l}_{\alpha \alpha}(\phi_{\alpha},x)$.
Here $V^{l}(\phi,\phi_{\alpha})$
is the effective inter-electron
potential of the Cooper pair in the $l$ side.
In our numerical calculations,
new $\Delta^{l}(\phi_{\alpha},x)$ and 
$\hat{g}^{l}_{\alpha \alpha}(\phi_{\alpha},x)$ are obtained
using Eqs.(\ref{req})-(\ref{gd++}) and (\ref{peq}).
We reiterate this process until the convergence is sufficiently
obtained.
\par
Based on the self-consistently determined
pair potentials, the LDOS can be calculated
as,
\begin{align}
N_{l}(E,x) &= \frac{1}{\pi}
{\displaystyle
\int ^{\pi/2}_{-\pi/2}d\phi
n_{l}(E,\phi,x)}, \\
\label{eq:LDOS}
n_{l}(E,\phi,x) &={\rm Im} \left \{
\frac{N_{0}}{2}
{\rm Tr}\left [
\hat{g}_{\alpha \alpha}^{l}
(\phi_{\alpha},x)\hat{\tau}_{3}
\right ] \right \}
_{{\rm i}\omega_m
\rightarrow
E+{\rm i}\delta},
\end{align}
where $N_{0}$ means the density of states (DOS)
in normal states, and $\delta$ is infinitesimal.
In this paper, 
we choose  the temperature $T$ as  $T/T_{d}=0.02$,
where $T_d$ is the critical temperature
of the bulk $d_{x^2-y^2}$-wave superconductor.
\par

\begin{figure}[htb]
\begin{center}
\scalebox{0.53}{
\includegraphics{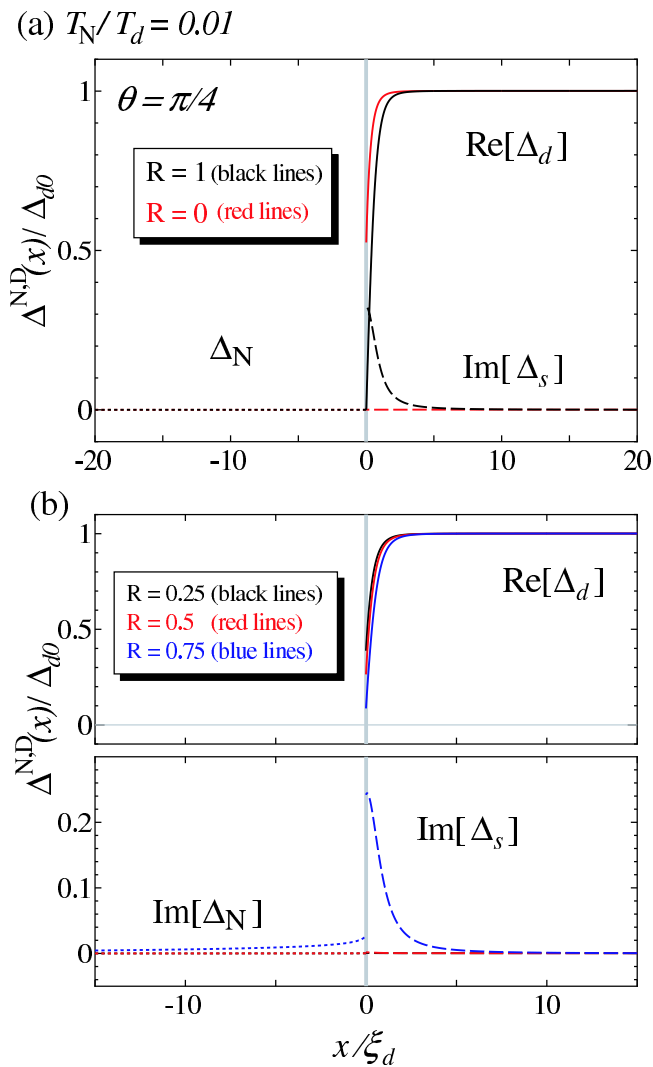}}
\caption{Spatial dependence of the pair potentials in 
N/D junctions with $\theta = \pi/4$ and
$T_{\rm N}/T_{d}=0.01$.
(a) $R=0$ and $1$,
(b) $R=0.25$, $0.5$, and $0.75$.
\label{fig:03}}
\end{center}
\end{figure}

\section{Results of numerical calculations}
\label{sec:03}
In this section, 
we show our results of numerical calculations on
the spatial dependence of the self-consistently determined
pair potentials and the corresponding LDOS.
The spatial variation of the 
pair potentials in the N/D junction
is expressed as
\begin{align}
\label{del:N}
\Delta^{\rm N}(\phi,x) &=\Delta_{\rm N}(x),
\\
\label{del:D}
\Delta^{\rm D}(\phi,x) &=\Delta_{d}(x)\cos2(\phi - \theta)
+\Delta_{s}(x),
\end{align}
where $\Delta_{\rm N}(x)$, $\Delta_{d}(x)$, and $\Delta_{s}(x)$
denote the amplitude of $s$-wave component in the N side,
$d_{x^2-y^2}$-wave component in the D side,
and subdominant $s$-wave component 
in the D side, respectively.
The attractive potentials $V^{l}(\phi,\phi^{\prime})$
with $l= {\rm N,D}$, 
are given by
\begin{align}
V^{\rm N}(\phi,\phi^{\prime}) &= V_{\rm N},
\\
V^{\rm D}(\phi,\phi^{\prime}) &= 2V_{d}
\cos (2\phi-2\theta)\cos (2\phi^{\prime}-2\theta) + V_{s},
\end{align}
and
\begin{align}
V_{\rm N} &= \frac{2\pi k_{\rm B}T}
{\displaystyle \ln \frac{T}{T_{\rm N}}
+ \sum_{0 \leq m < \omega_{c}/2\pi T}\frac{1}{m+1/2} },
\\
V_{d} &= \frac{2\pi k_{\rm B}T}
{\displaystyle \ln \frac{T}{T_{d}}
+ \sum_{0 \leq m < \omega_{c}/2\pi T}\frac{1}{m+1/2} },
\\
V_{s} &= \frac{2\pi k_{\rm B}T}
{\displaystyle \ln \frac{T}{T_{s}}
+ \sum_{0 \leq m < \omega_{c}/2\pi T}\frac{1}{m+1/2} }.
\end{align}
Here, $T_{\rm N}$ denotes the
transition temperature of $s$-wave pair potential 
in the N side.
$T_{s}$ denotes the transition temperature
of $s$-wave pair potential 
in the D side without $d_{x^2-y^2}$-wave
attractive potential.
\par
First,
in order to understand the role of
the proximity effect clearly, let us check the case of
normal metal/$s$-wave superconductor (N/S) junctions.
Figure~\ref{fig:02}(a) shows
the obtained spatial dependence of the pair potentials
in the junctions for extremely
low ($R=1$) and high ($R=0$) transparency cases.
The $x$-axis of Fig.~\ref{fig:02}
is normalized by $\xi_{s}=\hbar v_{\rm F}/\pi \Delta_{s0}$,
which is the coherence length of
$s$-wave superconductor.
For $R=1$,
the $s$-wave pair potential
remains to be constant in S side,
whereas the pair potential in the N side is zero.
The spatial variation of the pair potentials
is represented  as a step function.
The resulting LDOS at the interface
reproduces the bulk U-shaped DOS
[see Fig.~\ref{fig:02}(b)].
On the other hands, for $R=0$,
the pair potential $\Delta_{\rm N}$ in the N side
survives toward the inside.
We can see that the LDOS at the interface
in the presence of the proximity effect
is different from the bulk DOS.
\par
%

\begin{figure}[htb]
\begin{center}
\scalebox{0.65}{
\includegraphics{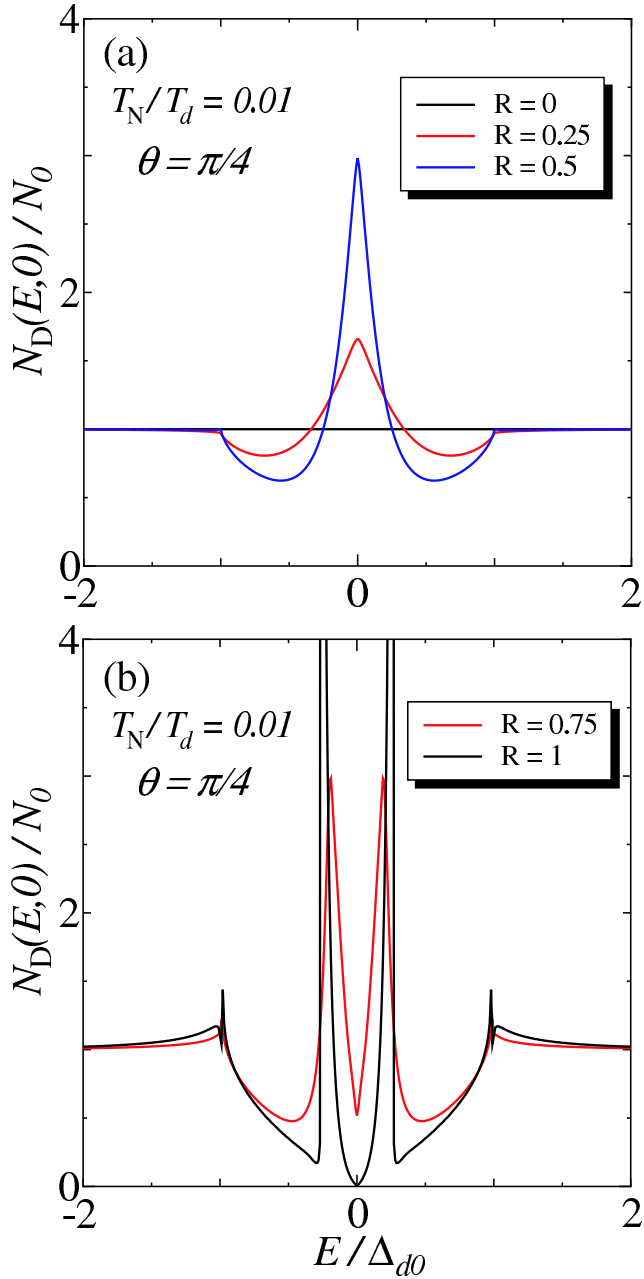}}
\caption{The LDOS for the N/D junction with $\theta = \pi/4$
and $T_{\rm N}/T_{d}=0.01$.
(a) $R=0$, $0.25$, and $0.5$,
(b) $R=0.75$ and $1$.
\label{fig:04}}
\end{center}
\end{figure}

\begin{figure}[htb]
\begin{center}
\scalebox{0.53}{
\includegraphics{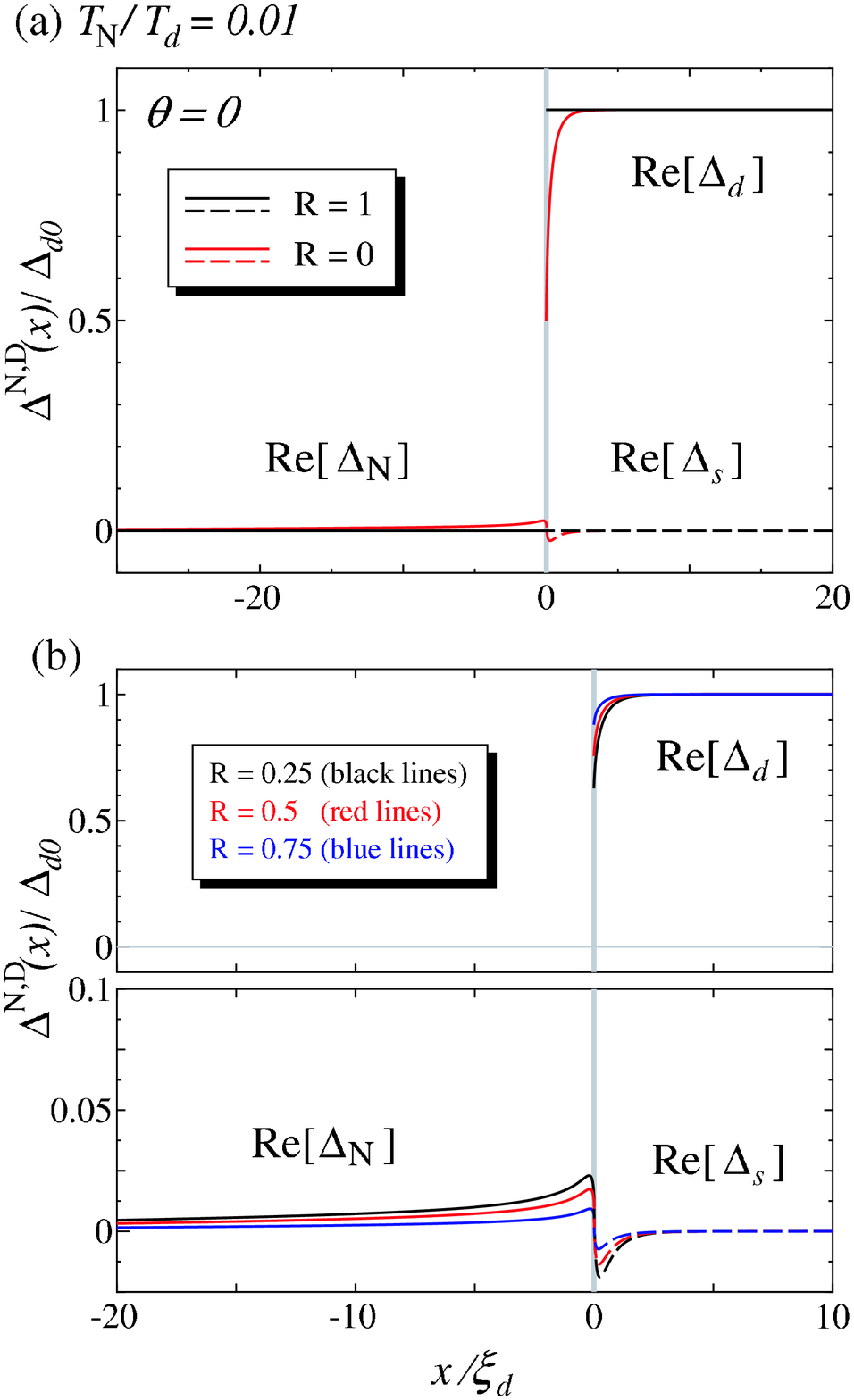}}
\caption{Spatial dependencies of the pair potentials 
in N/D junctions with $\theta = 0$ and
$T_{\rm N}/T_{d}=0.01$.
(a) $R=0$ and $1$,
(b) $R=0.25$, $0.5$, and $0.75$.
\label{fig:05}}
\end{center}
\end{figure}

\begin{figure}[htb]
\begin{center}
\scalebox{0.45}{
\includegraphics{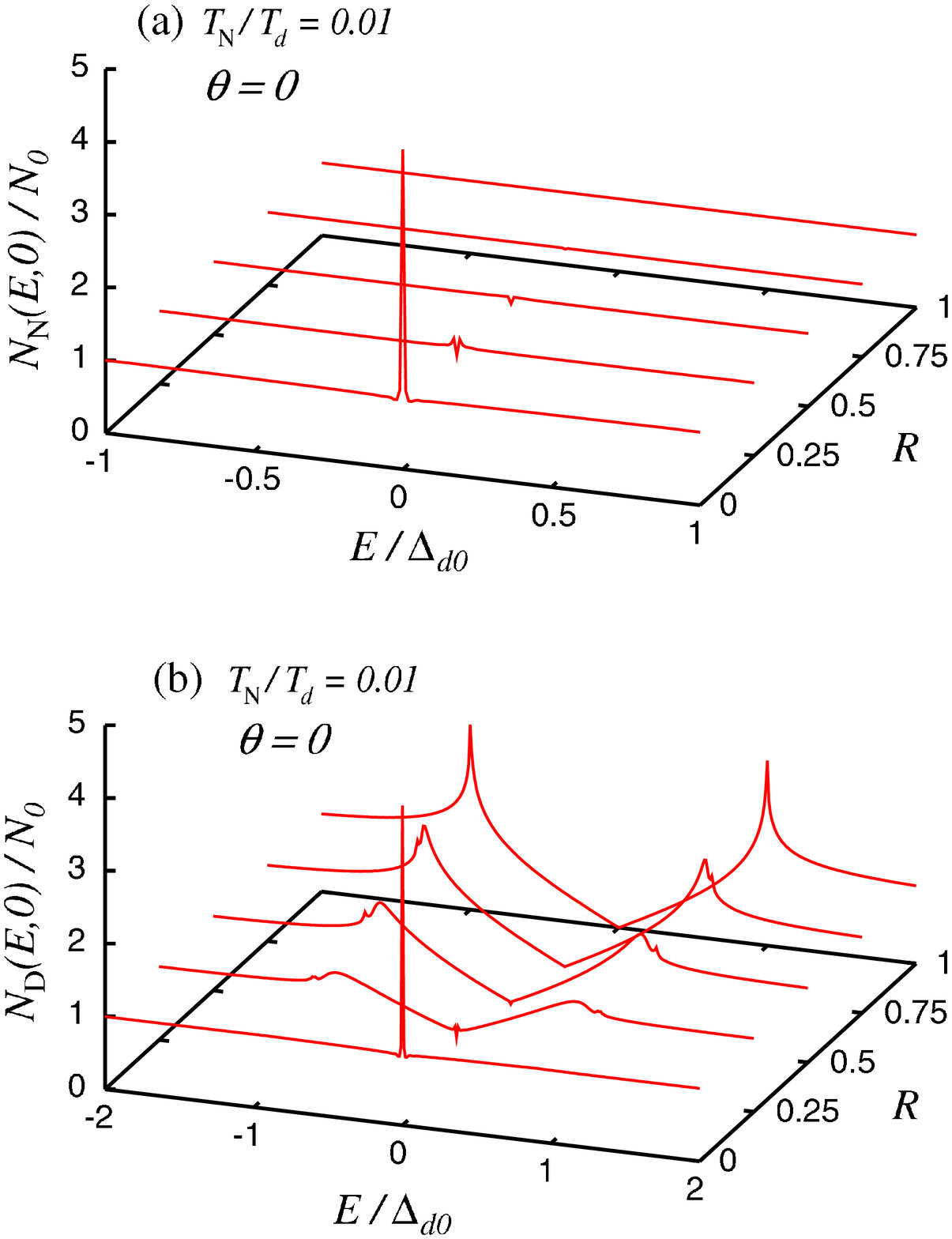}}
\caption{(a) The  LDOS at the interface 
by its value in the normal state for various $R$ 
with $\theta = 0$ and $T_{\rm N}/T_{d}=0.01$.
(a) The N side, and (b) the D side.
\label{fig:06}}
\end{center}
\end{figure}
%

\subsection{N/D junction with $\theta=\pi/4$}
\label{sec:03-1}
%
In this subsection, 
we focus on the N/D junction 
with (110)-oriented interface ($\theta = \pi/4$). 
As the parameter of the subdominant $s$-wave 
component, we take $T_{s}/T_{d}=0.3$ in the following. 
In Figs.~\ref{fig:03}(a) and \ref{fig:03}(b),
the spatial variations of the pair potentials
is shown for $T_{\rm N}/T_{d}=0.01$
for $R=0$, $1$, and $0.25$, $0.5$, $0.75$,
respectively.
Here,
Re[$\Delta_{d}(x)$] and Im[$\Delta_{{\rm N},s}(x)$]
denote real part of $\Delta_{d}(x)$ and imaginary
part of $\Delta_{{\rm N},s}(x)$, respectively.
The $x$-axis of Fig.~\ref{fig:03}
is normalized by the $d_{x^2-y^2}$-wave
coherence length $\xi_{d}=\hbar v_{\rm F}/\pi \Delta_{d0}$,
where $\Delta_{d0}$ is $d_{x^2-y^2}$-wave pair potential
in bulk states.
\par
First,
we present our results in the light of
previous theories. \cite{Hu,Tanaka9371,Matsu2}
The reduction of Re[$\Delta_{d}(x)$]
originates from a depairing effect that the
effective pair potentials $\Delta^{\rm D}(\phi_{+},0)$
and $\Delta^{\rm D}(\phi_{-},0)$ have reversed
contribution to the pairing interaction
for certain range of $\phi$.
Re[$\Delta_{d}(x)$] is suppressed at the interface
in the D side. 
At the same time, 
the quasiparticle forms the ABS with zero-energy
at the interface. \cite{Hu,Tanaka9371}
When the magnitude of the reflection probability
$R$ approaches unity, 
the ABS becomes unstable with the introduction of
the subdominant $s$-wave attractive potential in the D side.
And then Im[$\Delta_{s}(x)$] is induced
in the vicinity of the interface in the D side
in low transparent cases with large magnitude of $R$.
\cite{Matsu2}
As regards the proximity effect in the N side,
the pair potential $\Delta_{{\rm N}}(x)$
is not induced for fully high and low
transparent cases.
However,
the imaginary component of $\Delta_{{\rm N}}(x)$
is enhanced nearly in $R=0.75$
[see Fig.~\ref{fig:03}].
We can recognize that
the existence of Im[$\Delta_{s}(x)$]
near the interface of D side
can allow the enhancement of the Im[$\Delta_{{\rm N}}(x)$]. 
This fact is more recently found
by L\"{o}fwander, \cite{Lof}
and our results are consistent with his work.
\par
In Figs.~\ref{fig:04}(a) and \ref{fig:04}(b),
we show the corresponding LDOS
at the D side of the interface, 
where the same parameters of Figs.~\ref{fig:03}(a)
and \ref{fig:03}(b) are used, 
respectively.
For $R=0$, $N_{\rm D}(E,0)$ is equal to $N_{0}$
and has no $E$ dependence.
If we neglect the induced $s$-wave component in the N side,
the resulting LDOS has a zero-energy enhanced line shapes
for small $R$.
\cite{Tanu5}
This is because that the magnitude of the subdominant 
$s$-wave component, \textit{i.e.}, Im[$\Delta_{s}(x)$]
is very small and the incident and
reflected quasiparticles normal to the interface
can feel opposite signs of the $d_{x^2-y^2}$-wave
pair potentials.
However, as shown in Fig.~\ref{fig:03}(b),
not only Im[$\Delta_{{\rm N}}(x)$]
but also Im[$\Delta_{s}(x)$]
remain to be non-zero due to the proximity effect.
In this case, the resulting LDOS has complex 
line shapes.
For $R=1$, due to the enhancement of the magnitude of Im[$\Delta_{s}(x)$], 
the LDOS has the ZEP splitting, which is consistent with the 
previous works \cite{Tanu5,FRS97,Matsu2}. 
\par

\begin{figure}[htb]
\begin{center}
\scalebox{0.50}{
\includegraphics{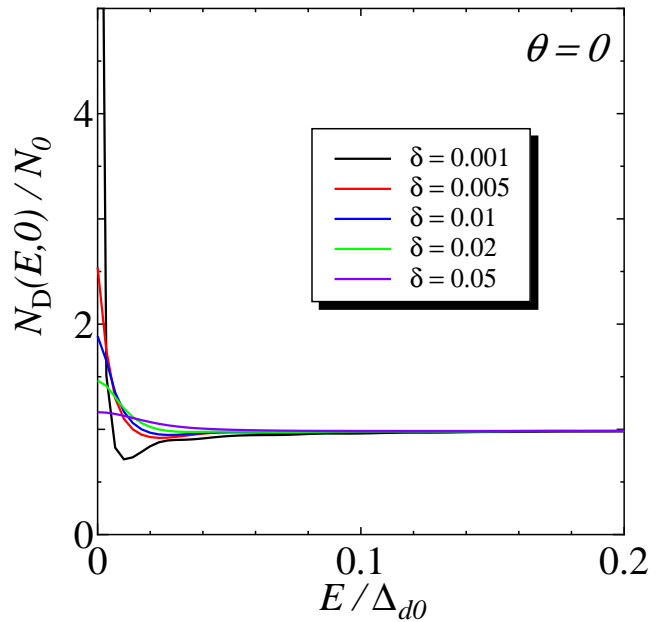}}
\caption{The LDOS at the interface near zero energy with $R=0$
for various $\delta$.
\label{fig:07}}
\end{center}
\end{figure}

\begin{figure}[htb]
\begin{center}
\scalebox{0.44}{
\includegraphics{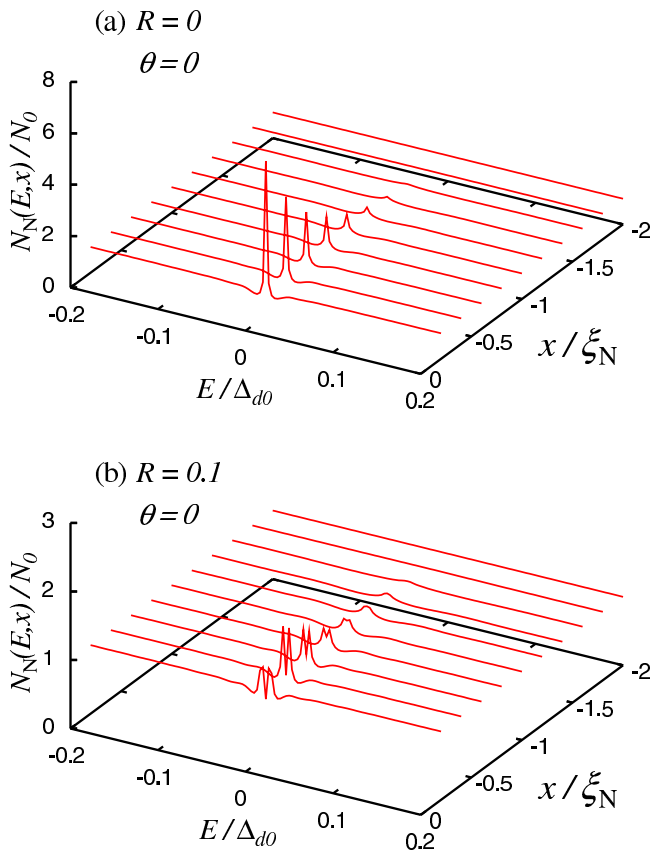}}
\caption{The LDOS at $x<0$ with $\theta = 0$.
(a) $R=0$ and (b) $R=0.1$ for various $x$.
Here
$\xi_{\rm N}=v_{\rm F}/2\pi T$ is the 
coherence length in the N side with $T=0.02T_{d}$. 
\label{fig:08}}
\end{center}
\end{figure}

\begin{figure}[htb]
\begin{center}
\scalebox{0.50}{
\includegraphics{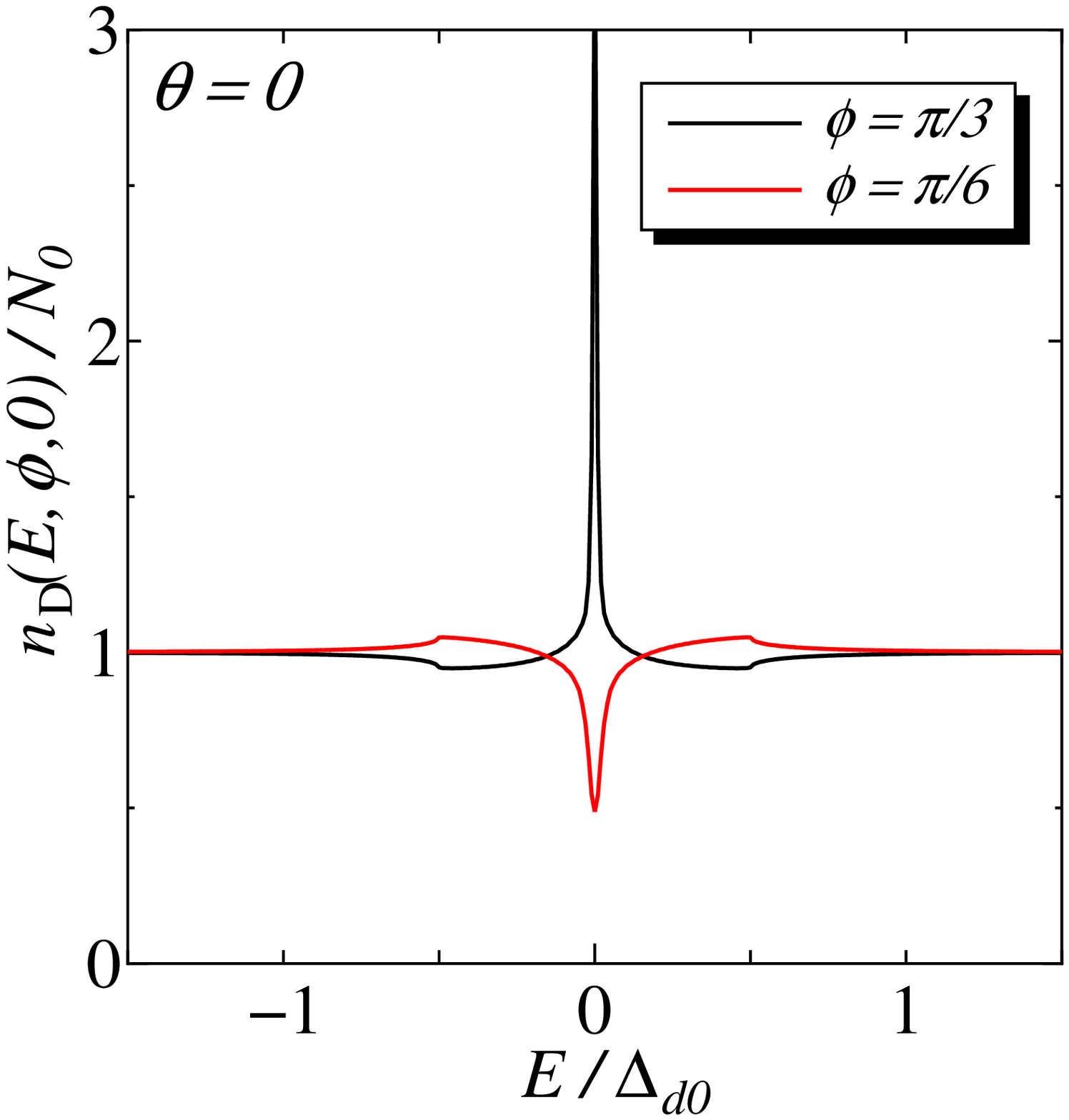}}
\caption{Angle resolved LDOS at the interface
for $\theta = 0$ and $R=0$.
\label{fig:09}}
\end{center}
\end{figure}

\subsection{N/D junction with $\theta=0$}
\label{sec:03-2}
In this subsection, 
we focus on the N/D junction with $\theta = 0$
and $T_{\rm N}/T_{d}=0.01$.
The spatial dependencies of the pair potentials
Re[$\Delta_{\rm N}(x)$], Re[$\Delta_{d}(x)$],
and Re[$\Delta_{s}(x)$],
are plotted in Figs.~\ref{fig:05}(a) and \ref{fig:05}(b),
for $R=0$, $1$, and $0.25$, $0.5$, $0.75$,
respectively.
For the low transparent limit, $R=1$, 
[see Fig.~\ref{fig:05}(a)],
the amplitude of the pair potential is constant.
This situation is similar to the case
of the N/S junctions with $R=1$
[see Fig.~\ref{fig:02}(a)].
With the decrease of $R$, 
Re[$\Delta_{d}(x)$] is suppressed near the interface,
while Re[$\Delta_{s}(x)$] is slightly mixed
in the vicinity of the interface.
Since Im[$\Delta_{s}(x)] =0$ is satisfied,
the TRS is not broken
near the (100) interface.
On the other hand for the N side,
we can readily see that the pair potential
Re[$\Delta_{\rm N}(x)$] is induced. 
This indicates that the superconducting
pair potential penetrates into the N side
due to the proximity effect.
\cite{Ohashi}
With increasing the magnitude of $T_{\rm N}$,
the amplitude of Re[$\Delta_{\rm N}(x)$] becomes
larger.
However, both Im[$\Delta_{s}(x)] =0$ and
Im[$\Delta_{\rm N}(x)] =0$ are satisfied. 
This situation is 
significantly different from the corresponding 
case of $\theta = \pi/4$ where the TRS is broken. 
\par
Next, let us look at the corresponding LDOS
at the interface. 
We show that the LDOS at the interface in 
the N and D sides 
as shown in Fig.~\ref{fig:06}.
For $R=1$, the LDOS on the D side has a V-shaped structure
similar to the bulk $d$-wave density of states.
\cite{Tanu5,FRS97,Matsu2}
The corresponding LDOS in the N side is constant. 
With the decrease of the magnitude of $R$, 
the magnitude of LDOS in the D side 
around zero energy is enhanced
while that around $\pm \Delta_{d0}$ is suppressed.
For $R=0.25$, the LDOS has a small dip like structure 
both in the N and D sides.
The fine structure around zero energy
is due to the induced 
pair potential $\Delta_{\rm N}(x)$ in the N side. 
The extreme case is $R=0$, where normal reflection 
is absent.  
The pair potential Re[$\Delta_{\rm N}(x)$]
induces the ZEP in the LDOS [see Fig.~\ref{fig:06}]
through Andreev reflection. \cite{Ohashi}
%
\par
Now we concentrate on the width of this ZEP.
We see that the relevance of the peak width 
and the infinitesimal number $\delta$ 
introduced in Eq.(\ref{eq:LDOS}) in order to avoid 
divergence in the actual calculation where 
the inverse of $\delta$ can be regarded
as a life time of quasiparticles. 
As shown in Fig.~\ref{fig:07}, the width of the ZEP becomes narrow
with the decrease of the magnitude of $\delta$, 
while the height of it increases monotonically. 
For $\delta \rightarrow 0$ limit, 
the ZEP is reduced to be expressed by the $\delta$-function.
\par
In Fig.~\ref{fig:08}, 
we show the spatial dependence of LDOS in the N side ($x<0$). 
For $R=0$, the zero energy states formed at the 
interface penetrate into the N side.
The width and the height of the ZEP is reduced 
with the increase of the magnitude of $x$. 
For $R=0.1$, the LDOS has a mini gap. 
The width of the gap has a spatial dependence
since the induced pair potential
Re[$\Delta_{\rm N}(x)$] depends on $x$.
The width of the mini gap is reduced  with the 
increase of the magnitude of $x$. 
\par
At the end of this subsection, 
let us remark on the difference
on the origin of ABS's between
$\theta =\pi/4$ and $\theta =0$.
As mentioned in Sec.~\ref{sec:03-1},
the formation of ABS's is originated from
the opposite sign of the pair potentials 
between $\Delta_{d}(\phi_{+},0)$ and
$\Delta_{d}(\phi_{-},0)$ felt by 
quasiparticles in the D side.
However, in the case of $\theta =0$,
there is no sign change between 
$\Delta_{d}(\phi_{+},0)$ and
$\Delta_{d}(\phi_{-},0)$ for any $\phi $.
This manifestation of the ZEP in the LDOS
is due to the another origin. 
It is due to the sign change of the pair potentials 
between Re$[\Delta_{\rm N}(x)]$
and Re$[\Delta_{s}(x)]$
or that between Re$[\Delta_{\rm N}(x)]$ 
and Re$[\Delta_{d}(x)]$ for $|\phi |>\pi/4$
through the Andreev reflection.
Actually,
by calculating
the angle resolved LDOS $n_{\rm D}(E,\phi,x)$
in Eq.(\ref{eq:LDOS}), 
$n_{\rm D}(E,\phi,x)$ has the ZEP for $\phi =\pi/3$
while it does not for $\phi =\pi/6$
as shown in Fig.~\ref{fig:09}.
\par
%

\begin{figure*}[htb]
\begin{center}
\scalebox{0.90}{
\includegraphics{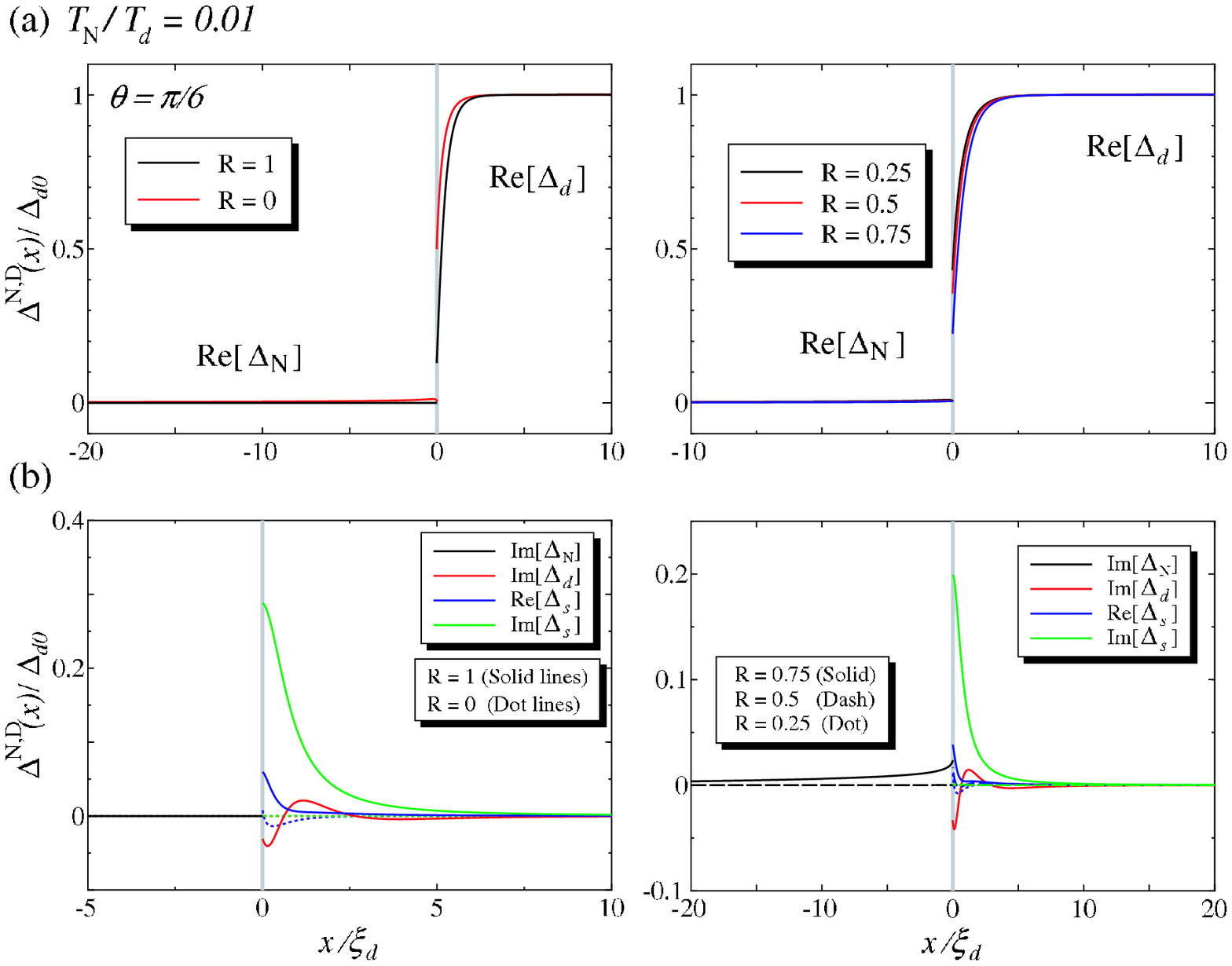}}
\caption{Spatial dependencies of the pair potentials  for 
the N/D junctions with $\theta = \pi/6$
and $T_{\rm N}/T_{d}=0.01$.
In (a), Re$[\Delta_{\rm N}(x)]$,
Re$[\Delta_{d}(x)]$, and
in (b), Im$[\Delta_{\rm N}(x)]$, Im$[\Delta_{d}(x)]$,
Re$[\Delta_{s}(x)]$, Im$[\Delta_{s}(x)]$,
are plotted, respectively.
\label{fig:10}}
\end{center}
\end{figure*}
%

\begin{figure}[htb]
\begin{center}
\scalebox{0.65}{
\includegraphics{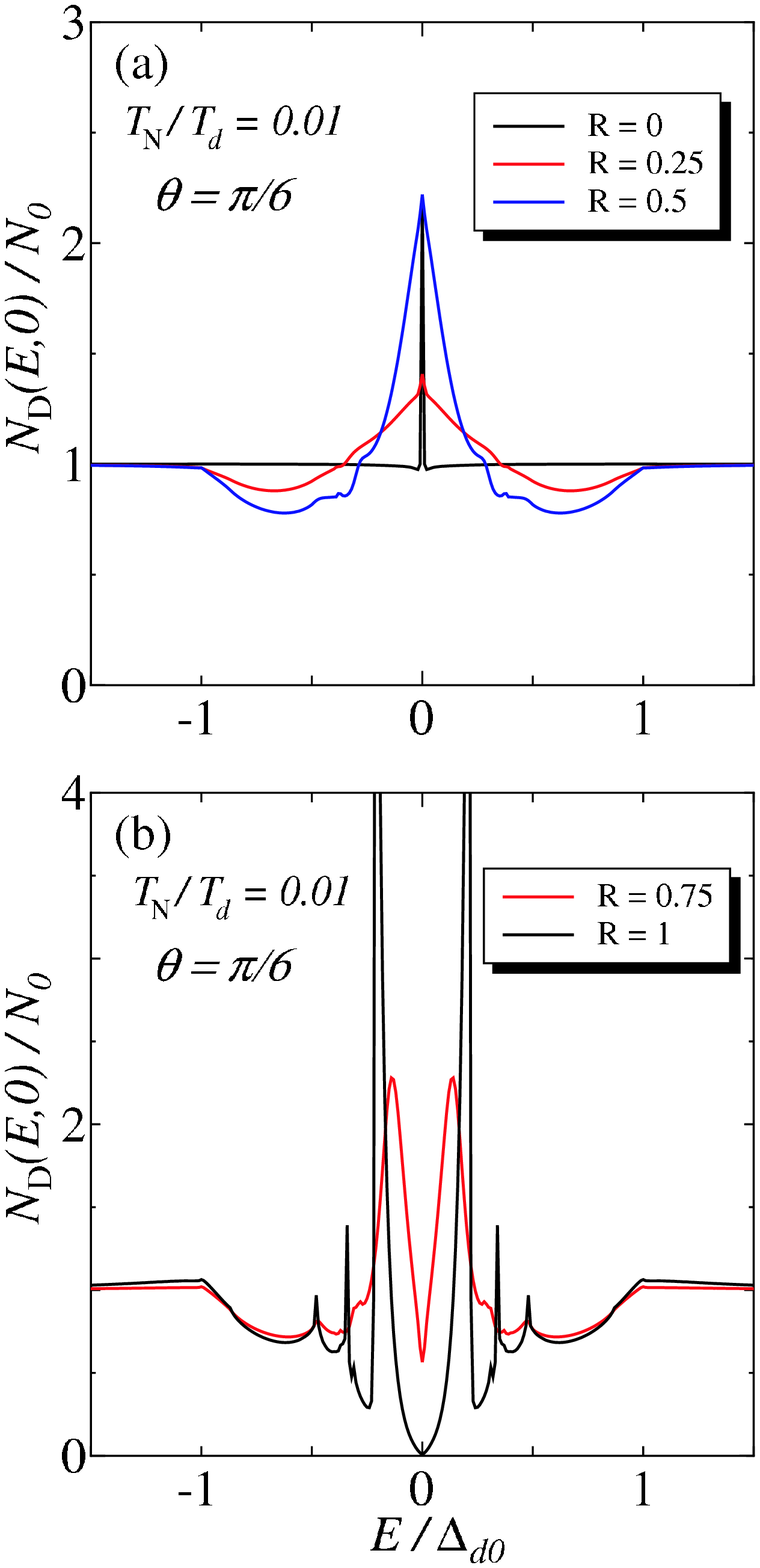}}
\caption{The LDOS for the N/D junctions
with $\theta = \pi/6$ and $T_{\rm N}/T_{d}=0.01$.
(a) $R=0$, $0.25$, and $0.5$,
(b) $R=0.75$ and $1$.
\label{fig:11}}
\end{center}
\end{figure}
\subsection{N/D junction with $\theta=\pi/6$}
\label{sec:03-3}
In this subsection, we concentrate on the case
with intermediate angle $\theta$,
\textit{e.g.}, $\theta = \pi/6$.
The spatial dependencies of the pair potentials 
become very complicated as shown in  Fig.~\ref{fig:10},
where we choose $T_{\rm N}/T_{d}=0.01$ for various $R$. 
For $R=0$, 
the amplitude of Re[$\Delta_{\rm N}(x)$] is induced 
toward the inside of N.
This situation is similar to the case of $\theta =0$
by the proximity effect.
With the increase of $R$,
the amplitude of Re[$\Delta_{d}(x)$] is significantly
suppressed at the interface, 
due to the destructive interference originating
from the sign changing 
nature of $d_{x^{2}-y^{2}}$-wave pair potential. 
The amplitude of Re[$\Delta_{\rm N}(x)$] is suppressed 
due to the reduction of the proximity effect. 
On the other hand, amplitudes of Im[$\Delta_{d}(x)$], 
Re[$\Delta_{s}(x)$], and Im[$\Delta_{s}(x)$] are enhanced. 
%
As regards the subdominant components,
the amplitude of Im[$\Delta_{s}(x)$] for lower transparent
cases $R=0.75$ and $R=1$,
is one order larger than those of
Im[$\Delta_{d}(x)$], and Re[$\Delta_{s}(x)$]
[see Fig.~\ref{fig:10}(b)].
In particular, for $R=0.75$, the situation that
the amplitude of Im[$\Delta_{\rm N}(x)$] is induced near
the interface of N side
is similar to the case of $\theta =\pi/4$.
Subdominant components which break TRS
is induced also by the proximity effect
and depends on the transparencies of the junction.
\par
The corresponding LDOS for $\theta = \pi/6$
are plotted in Figs.~\ref{fig:11}(a)
and \ref{fig:11}(b).
Here, we choose $T_{\rm N}/T_{d}=0.01$,
which is the same parameters
used in Fig.~\ref{fig:10}.
The resulting line shapes of LDOS
are complex reflecting on the complicated
spatial dependence of the pair potentials.
For $R=0$,
we can see that the zero energy states
are formed at the interface
and the resulting LDOS has a ZEP 
due to the \textit{proximity effect}. 
This ZEP, the origin of which is similar
to the case of $\theta =0$ [see Fig.~\ref{fig:06}(b)],
arises from the sign change of the pair potentials 
between Re[$\Delta_{\rm N}(x)$] and Re[$\Delta_{d,s}(x)$]
through Andreev reflections due to proximity effect.
With the increase of $R$, the LDOS has a rather broad ZEP.
due to the formation of \textit{ABS's}.
This ZEP for intermediate value of $R$ originates from
the formation of ABS's peculiar to unconventional
superconductors \cite{TK95,KT00},
\textit{i.e.} with sign change of the pair potentials
between $\Delta_{d}(\phi_{+},0)$ and $\Delta_{d}(\phi_{-},0)$.
With the further increase of $R$, (see the curve with  $R=0.75$),
since the imaginary components of both $\Delta_{\rm N}(x)$
and $\Delta_{s}(x)$ are induced in the vicinity of the interface,
the line shape of the LDOS has a complicated structure
reflecting on the complex spatial dependence of pair potentials.
For $R=1$, the LDOS has the ZEP splitting
due to the existence of subdominant
$s$-wave component which breaks TRS in the D side,
\textit{i.e.}, Im[$\Delta_{s}(x)$].
As a results,
in the N/D junctions with $\theta = \pi/6$,
we can conclude that the line shapes of the LDOS
around zero energy
change from  
(I) \textit{the ZEP due to proximity effect},  
(II) \textit{the ZEP due to the ABS},
and (III) \textit{the ZEP splitting due to the formation of the BTRSS}, 
with the increase of the magnitude of $R$.
\par
\begin{figure*}[htb]
\begin{center}
\scalebox{0.88}{
\includegraphics{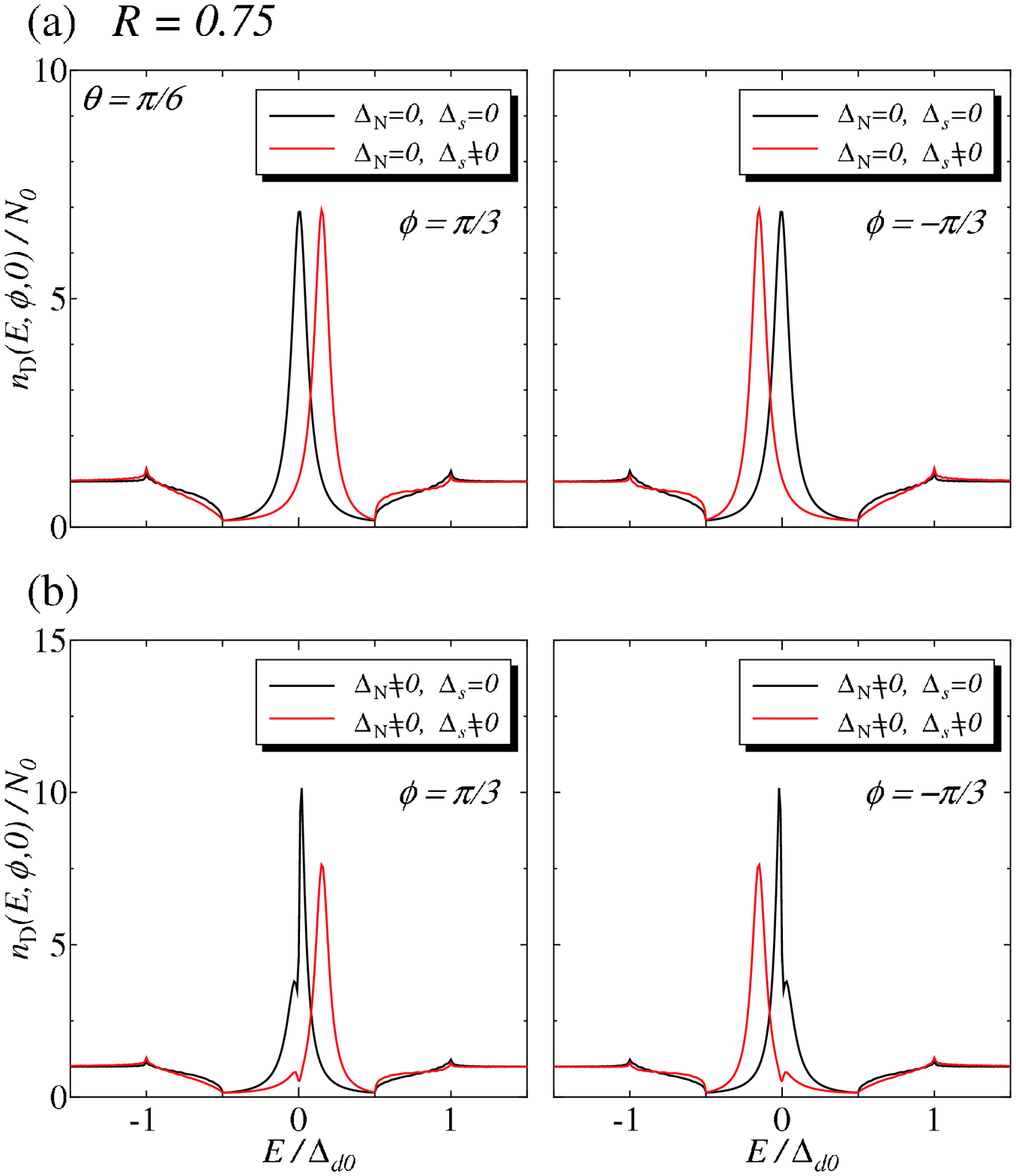}}
\caption{The angle resolved LDOS $n_{\rm D}(E,\phi,x)$
at the interface with $\theta = \pi/6$ and $R=0.75$.
(a): LDOS without subdominant $s$-wave pair potential in the N side.
(b): LDOS in the presence of subdominant $s$-wave pair potential
in the N side.
$\phi = \pi/3$ for the left panel and 
$\phi = -\pi/3$ for the right panel.
\label{fig:12}}
\end{center}
\end{figure*}
\begin{figure*}[htb]
\begin{center}
\scalebox{0.88}{
\includegraphics{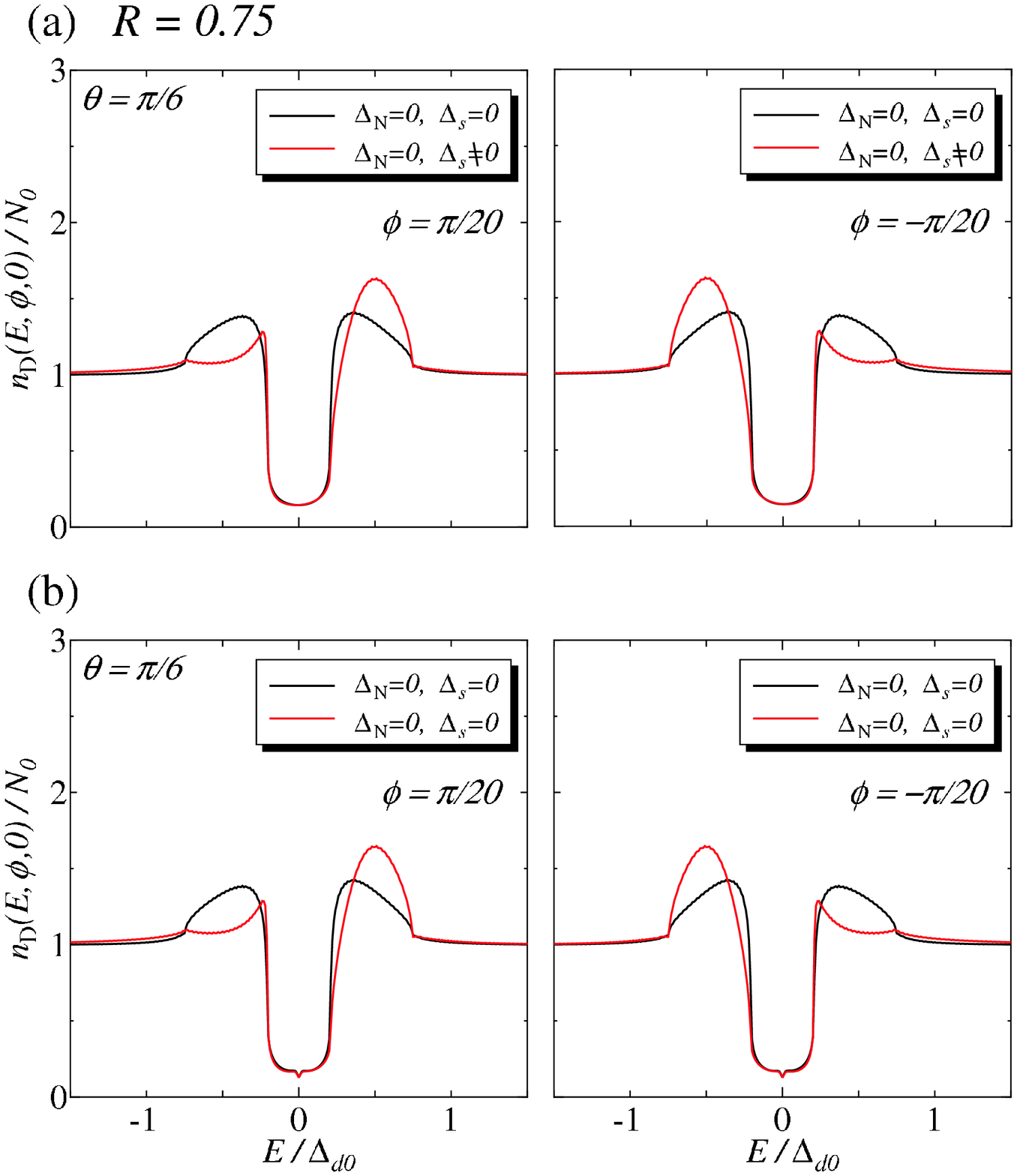}}
\caption{The angle resolved LDOS $n_{\rm D}(E,\phi,x)$
at the interface with $\theta = \pi/6$ and $R=0.75$.
(a): LDOS without subdominant $s$-wave pair potential in the N side.
(b): LDOS in the presence of subdominant $s$-wave pair potential
in the N side. $\phi = \pi/20$ for the left panel and 
$\phi = -\pi/20$ for the right panel.
\label{fig:13}}
\end{center}
\end{figure*}

Finally,
we look at the angle resolved LDOS
$n_{\rm D}(E,\phi,x)$ in order to understand the basic features
of the line shapes of $N_{\rm D}(E,0)$.
In Figs.~\ref{fig:12} and \ref{fig:13},
the angle resolved LDOS $n_{\rm D}(E,\phi,x)$
at the interface is plotted for $\theta = \pi/6$ and $R=0.75$.
In order to understand the role of the induced pair potential 
in N region,  we intentionally neglect the pair potential
$\Delta_{\rm N}(x)$ in the N side in the actual calculation of LDOS 
as shown in the upper panel of Fig.~\ref{fig:12} and Fig.~\ref{fig:13}. 
We also consider the case, where induced $s$-wave component 
$\Delta_{s}(x)$ is neglected in the calculation of LDOS.
First we concentrate on the upper panels
where $\Delta_{N}(x)$ is absent.
If only the pure $d_{x^2-y^2}$-wave component exists
in the D side, 
the quasiparticles form the ABS's at zero energy.
For $\phi = \pm \pi/3$,
since $\Delta_{d}(\phi_{+},0) \Delta_{d}(\phi_{-},0)<0$
is satisfied, $n_{\rm D}(E,\phi,0)$ has the ZEP
[see Figs.~\ref{fig:12}(a) and \ref{fig:12}(b)].
In the presence of the subdominant $s$-wave component 
which breaks TRS in the D side,
the resulting $n_{\rm D}(E,\phi,0)$ is modified.
In this case, the line shape of $n_{\rm D}(E,\phi,0)$
is not asymmetric around $E=0$, \textit{i.e.},
where $n_{\rm D}(E,\phi,0) \neq n_{\rm D}(-E,\phi,0)$, 
due to the formation of the BTRSS.
Therefore, the peak position of $n_{\rm D}(E,\phi,0)$
is shifted from zero energy.
On the other hands, for $\phi = \pm \pi/20$,
as $\Delta_{d}(\phi_{+},0) \Delta_{d}(\phi_{-},0)>0$
is satisfied, $n_{\rm D}(E,\phi,0)$ has gap structures
[see Figs.~\ref{fig:13}(a) and \ref{fig:13}(b)].
These results are consistent with those
by Matsumoto and Shiba.
\cite{Matsu3}
Next, we look at lower panels of  Figs.~\ref{fig:12} and \ref{fig:13}. 
Since  induced Im[$\Delta_{\rm N}(x)$] breaks time reversal symmetry, 
the resulting line shape of LDOS is no more symmetric around $E=0$, 
even for $\Delta_{s}(x)=0$. 
As shown in the lower panels, the presence of $\Delta_{\rm N}(x)$
forms a minigap around zero energy $E=0$.
\par
\begin{figure}[htb]
\begin{center}
\scalebox{0.65}{
\includegraphics{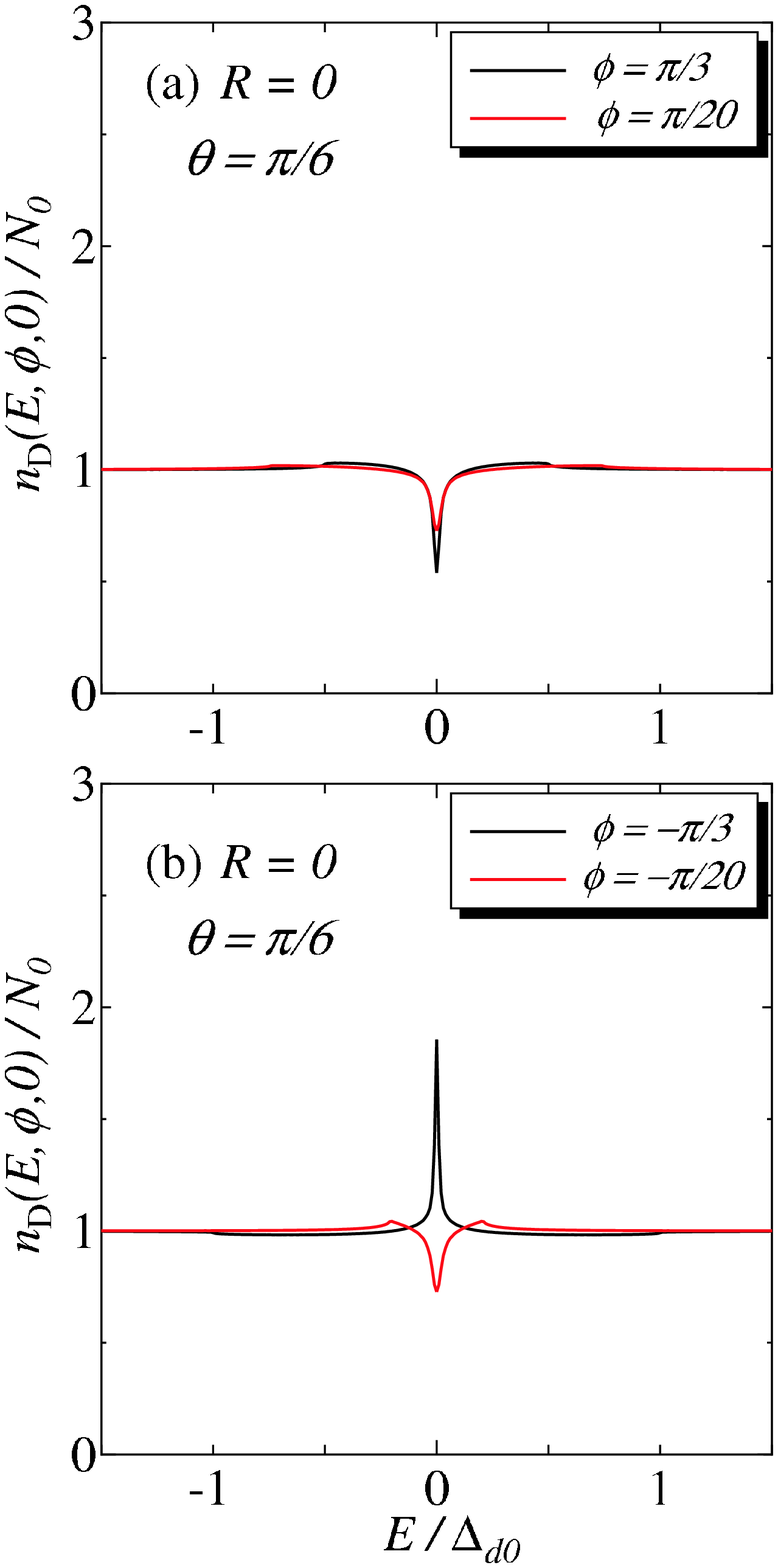}}
\caption{The angle resolved LDOS $n_{\rm D}(E,\phi,x)$
at the interface with $\theta = \pi/6$ and $R=0$.
(a) $\phi = \pi/3$, $\pi/20$, and
(b) $\phi =-\pi/3$, $-\pi/20$. 
\label{fig:14}}
\end{center}
\end{figure}

In Fig.~\ref{fig:14},
the LDOS $n_{\rm D}(E,\phi,0)$ is plotted
for $R=0$, where the Andreev reflection only exists. 
In this case, the line shape of the LDOS can be understood only 
by taking into account the pair potentials
$\Delta_{\rm N}(x)$ and $\Delta_{d}(x)$. 
For $\phi = \pi/3$ and $\phi = \pm \pi/20$,
the LDOS has no ZEP since
$\Delta_{\rm N}(\phi_{+},0) \Delta_{d}(\phi_{+},0)>0$
is satisfied. 
The resulting  $n_{\rm D}(E,\phi,0)$ has a gap structure
[see Figs.~\ref{fig:14}(a) and \ref{fig:14}(b)].
If the quasiparticle's scattering
feel different sign of the pair potentials
between the N and D sides,
we can expect that the $n_{\rm D}(E,\phi,0)$ has the ZEP even for $R=0$.
For $\phi = -\pi/3$, since 
$\Delta_{\rm N}(\phi_{+},0) \Delta_{d}(\phi_{+},0)<0$
is satisfied, then the  $n_{\rm D}(E,\phi,0)$
has the ZEP [see Fig.~\ref{fig:14}(b)].
These situations are similar to those studied in Sec.~\ref{sec:03-2}.
%
%
%
\par

\section{Conclusions}
\label{sec:04}
%
In this paper, 
we have studied quasiparticle properties in the N/D junctions,
in the presence of the proximity effect
as well as the BTRSS in the D side. 
We assume the attractive interelectron potentials 
which induces subdominant $s$-wave components
both in the N and D sides.  
The spatial dependencies of the pair potentials 
in the N/D junctions are determined self-consistently. 
The LDOS at the interface of the N/D junctions 
are studied in detail by changing orientational angle $\theta$
and the transparency of the junction. 
Our main results are summarized as follows.
\par
i)
For (110) oriented junction with $\theta=\pi/4$, 
the predominant $d_{x^2-y^2}$-wave component in the D side
is reduced at the interface. 
The subdominant $s$-wave component which breaks TRS
appears near the (110) interface in low transparent cases.
In the N side, subdominant imaginary $s$-wave component 
is induced near the interface \cite{Lof}. 
The resulting LDOS at the interface in the N side 
has the ZEP or its splitting
depending on the transparency of the junction.
These results are consistent with previous theoretical
works. \cite{Tanu5,FRS97,Lof,Matsu2,Matsu3}
\par
%
ii) For (100) oriented junction, 
the subdominant $s$-wave component becomes 
real number and does not break TRS \cite{Lof}. 
With the increase of the transparency of the junction, 
the magnitude of the $s$-wave component in the N side
is enhanced by the proximity effect. 
For fully transparent case, 
the LDOS has very sharp ZEP due to the formation of 
zero energy states.
This ZEP originates from the fact that
quasiparticles feel different sign
of the pair potentials between
the N and D sides through Andreev reflection.
\par
iii) For $\theta = \pi/6$,
the spatial dependencies of the pair potentials
become very complicated. 
The resulting LDOS has a wide variety.
For high barrier limit, the LDOS has the ZEP splitting.
\par
In the light of our results,
although the penetration of the pair potential into the N side
by the proximity effect 
is expected for (100) oriented junctions, 
the subdominant $s$-wave components both in 
the N and D sides are real numbers. 
Thus, the BTRSS is not formed in the (100) junctions.
The present results are inconsistent with 
previous prediction based on 
tunneling experiment on high-$T_{c}$ cuprates
by Kohen \textit{et al.}
\cite{Kohen}
In order to understand their experimental results,
we must seek for other possibilities. 
\par
In the present paper, we study proximity effect in 
N/D junctions in the ballistic limit.
In the actual junctions, we can not neglect 
impurity scattering effect.
To reply this issue, two of the present author 
present theoretical works of charge transport in 
N/D junctions in the presence of the ABS 
where the N region is a diffusive metal. 
\cite{r2,Tanaka2004}
It is an interesting problem to extend the present theory 
in the diffusive regime based on the Keldysh Green's function
formalism.
\cite{Tanaka2004}
Although there are many works about ABS's in unconventional superconductor
junctions up to now, \cite
{a,b,c,d,e,f,g,h,i,j,k,l,m,n,o,p,q,r,s,t,u,v,w,x,o1,o2,o3,o4,o5,K95,m1,m2,m3}
proximity effects both in the presence of induced pair potential
in the N region and the ABS's are not fully studied. 
Recent study by L\"{o}fwander is remarkable 
where the proximity effect is studied in the 
presence of both impurity scattering
and induced subdominant component of the 
pair potentials. \cite{Lof}
Based on his detailed calculation,  
for (100) junction, a real combination of the proximity effect 
is always found. 
For (110) orientation, the $s$-wave component induced
by the proximity effect in the N side breaks TRS. 
\cite{r4,Lof}
These results are consistent with the present results. 
\par
As a future problem, to clarify the charge transport property
in Josephson junctions both in the presence of ABS
\cite{J1,J2,J3,J4,J5,J6,J7,J8,J9,J10,J11,J12,J13,J14,J15,J16,J17,J18,J19}
and proximity effect is interesting.
We want to extend the present theory towards these directions. 
\par
\section*{Acknowledgments}
We would like to sincerely thank Dr. T. L\"{o}fwander
for critical reading of the manuscript 
and valuable discussions.
The computations have been performed
at the Supercomputer Center of
Yukawa Institute for Theoretical Physics,
Kyoto University.

\appendix

\section{Bogoliubov-de Gennes equation
and quasi-classical approximation}
\label{app:01}
Our theoretical formalism is started from
the Bogoliubov-de Gennes (BdG) equation
for unconventional spin-singlet superconductors,
\begin{align}
E_{n}\tilde{u}_{n}(\bm{r}) &= H_{0}\tilde{u}_{n}(\bm{r})
+ \int d \bm{r^{\prime}}
\Delta(\bm{r},\bm{r^{\prime}})
\tilde{v}_{n}(\bm{r^{\prime}}),
\\
E_{n}\tilde{v}_{n}(\bm{r}) &= -H_{0}\tilde{v}_{n}(\bm{r})
+ \int d \bm{r^{\prime}}
\Delta(\bm{r},\bm{r^{\prime}})^{*}
\tilde{u}_{n}(\bm{r^{\prime}}), \\
\nonumber
H_{0} &= -\frac{\hbar^{2}}{2m}\nabla^{2} - \mu,
\end{align}
where $\mu$ is the chemical potential,
while $\tilde{u}_{n}(\bm{r})$ and $\tilde{v}_{n}(\bm{r})$
denote the electron like and holelike components
of the wave function
\begin{align}
\tilde{\Psi}_{n}(\bm{r}) &=
\left (
    \begin{array}{c}
        \tilde{u}_{n}(\bm{r}) \\
        \tilde{v}_{n}(\bm{r})
    \end{array}
\right ), \nonumber \\
&\equiv
\left (
    \begin{array}{c}
        u_{n}(\hat{\bm{k}},\bm{r}) \\
        v_{n}(\hat{\bm{k}},\bm{r})
    \end{array}
\right ) e^{{\rm i}k_{\rm F} \hat{\bm{k}} \cdot \bm{r}}
=\Psi_{n}(\hat{\bm{k}},\bm{r})
e^{{\rm i}k_{\rm F} \hat{\bm{k}} \cdot \bm{r}}.
\end{align}
Here the quantities
$\hat{\bm{k}}$ and $\bm{r}$
stand for the unit vector of
the wave number of the Cooper pair which
is fixed on the Fermi surface
$(\hat{\bm{k}}=\bm{k}_{\rm F}/|\bm{k}_{\rm F}|)$,
and the  position of the center of mass of Cooper pair,
respectively. 
After applying the quasi-classical approximation,
the BdG equation is reduced to the Andreev equation
\cite{Hu,Bruder,Kurki},
\begin{align}
E_{n}\Psi_{n}(\hat{\bm{k}},\bm{r})
   &= -\left [ {\rm i}\hbar v_{\rm F}\hat{\bm{k}} \cdot \nabla
      + \hat{\Delta}(\hat{\bm{k}},\bm{r}) \right ]
       \hat{\tau}_{3}\Psi_{n}(\hat{\bm{k}},\bm{r}), \\
\hat{\Delta}(\hat{\bm{k}},\bm{r}) &=
\left (
    \begin{array}{cc}
        0 & \Delta(\hat{\bm{k}},\bm{r}) \\
  -\Delta(\hat{\bm{k}},\bm{r})^{*} & 0
    \end{array}
\right ).
\end{align}
The wave function $\Psi_{n}(\hat{\bm{k}},\bm{r})$
is obtained by neglecting the rapidly oscillating
plane-wave part following the quasi-classical approximation
\cite{Bruder,Kurki}.
The $\hat{\bm{k}}$ dependence of
$\Delta(\hat{\bm{k}},\bm{r})$
represents the symmetry of the pair potential.
\par
In the present study, we consider the case where a secularly
reflecting surface or interface runs along the $y$ direction.
Then, 
the pair potential depends only on $x$
since the system is homogeneous along the $y$ direction.
The wave function $\Psi_{n}(\hat{\bm{k}},\bm{r})$ 
can be written in the following directional
notation:\cite{Matsu2,Ashida}
\begin{align}
\Psi_{n}(\hat{\bm{k}},x,y) =& \left [
 \Phi_{n}^{(+)}(\phi_{+},x)e^{{\rm i}|k_{{\rm F}x}|x}
\right .
\nonumber \\
&+ \left .
\Phi_{n}^{(-)}(\phi_{-},x)e^{-{\rm i}|k_{{\rm F}x}|x}
\right ] e^{{\rm i}|k_{{\rm F}y}|y},
\nonumber \\
\Phi_{n}^{(\alpha)}(\phi_{\alpha},x) =&
\left (
    \begin{array}{c}
        u_{n}^{(\alpha)}(\phi_{\alpha},x) \\
        v_{n}^{(\alpha)}(\phi_{\alpha},x)
    \end{array}
\right ).
\end{align}
Here $\pm$ represents
the sign of the $x$ component of the Fermi wave number
$k_{{\rm F}x}$ and $\alpha (\beta) = \pm$.
We define a Green's function
$G_{\alpha \beta}(\phi,x,x^{\prime})$
and a quasi-classical Green's function
$g_{\alpha \beta}(\phi,x)$
\begin{align}
G_{\alpha \beta}(\phi,x,x^{\prime})
&= \sum_{n}\frac{\Phi_{n}^{(\alpha)}(\phi_{\alpha},x)
\Phi_{n}^{(\beta)}(\phi_{\beta},x^{\prime})^{\dagger}}
{{\rm i}\omega_{m} -E_{n}}, \\
g_{\alpha \beta}(\phi,x) \pm {\rm i}
(\hat{\gamma}_{3})_{\alpha \beta}
&= -2\hbar|v_{{\rm F}x}|\hat{\tau}_{3}
G_{\alpha \beta}(\phi,x\pm 0,x).
\end{align}
where $\hat{\gamma}_{3}$ is the Pauli matrix
in the directional space\cite{Ashida}.
\par

\section{Evolution operator
$\tilde{U}^{l}_{\alpha}(\phi_{\alpha},x,x^{\prime})$}
\label{app:02}
The quasi-classical Green's function
can be written by the following evolution operator
$U_{\alpha}(\phi_{\alpha},x,x^{\prime})$ as
\begin{eqnarray}
g_{\alpha \beta}(\phi,x)
   =U_{\alpha}(\phi_{\alpha},x,x^{\prime})
     g_{\alpha \beta}(\phi,x^{\prime})
      U_{\beta}^{-1}(\phi_{\beta},x,x^{\prime}),
\end{eqnarray}
where $U_{\alpha}(\phi_{\alpha},x,x^{\prime})$
satisfies the Andreev equation
\begin{align}
{\rm i}\hbar |v_{{\rm F}x}| &
\frac{\partial}{\partial x}
U_{\alpha}(\phi_{\alpha},x,x^{\prime})
\nonumber \\
&=-\alpha
\left [ {\rm i}\omega_{m}\hat{\tau}_{3}
+\hat{\Delta}(\phi_{\alpha},x)\right ]
U_{\alpha}(\phi_{\alpha},x,x^{\prime}),
\end{align}
with $U_{\alpha}(\phi_{\alpha},x,x)=1$.
\par
Hence,
the evolution operators in the $l$ side
can be divided
into a growing part and a decaying part
\begin{align}
U_{\alpha}^{l}(\phi_{\alpha},x,x^{\prime}) =&
\Lambda_{\alpha}^{l(+)}(\phi_{\alpha},x,x^{\prime})
e^{{\kappa^{l}_{\alpha}}(x-x^{\prime})}
\nonumber \\
&+ \Lambda_{\alpha}^{l(-)}(\phi_{\alpha},x,x^{\prime})
e^{-{\kappa^{l}_{\alpha}}(x-x^{\prime})},
\\
\Lambda_{\alpha}^{l(+)}(\phi_{\alpha},x,x^{\prime})
=& -\frac{1}{W^{l}_{\alpha}}
\Phi_{n}^{l(+)}(\phi_{\alpha},x)
{}^{\rm T}\Phi_{n}^{l(-)}(\phi_{\alpha},x^{\prime})
\hat{\tau}_{2},
\nonumber \\
\Lambda_{\alpha}^{l(-)}(\phi_{\alpha},x,x^{\prime})
=& \frac{1}{W^{l}_{\alpha}}
\Phi_{n}^{l(-)}(\phi_{\alpha},x)
{}^{\rm T}\Phi_{n}^{l(+)}(\phi_{\alpha},x^{\prime})
\hat{\tau}_{2},
\end{align}
where
\begin{align}
\kappa^{l}_{\alpha} &=
\frac{\Omega^{l}_{\alpha}}{|v_{{\rm F}x}|},
\quad
\Omega_{\alpha}^{l}=\sqrt{\omega_{m}^{2}
+|\Delta^{l}(\phi_{\alpha},\infty)|^{2}},
\\
W^{l}_{\alpha} &=
{}^{\rm T}\Phi_{n}^{l(+)}(\phi_{\alpha},x)\hat{\tau}_{2}
\Phi_{n}^{l(-)}(\phi_{\alpha},x)
\nonumber \\
&=-{}^{\rm T}\Phi_{n}^{l(-)}(\phi_{\alpha},x)\hat{\tau}_{2}
\Phi_{n}^{l(+)}(\phi_{\alpha},x)
\nonumber \\
&= \mathrm{const.}
\end{align}
In the above, $^{\rm T}\Phi_{n}^{l(\alpha)}(\phi_{\alpha},x)$
denotes the transposition of
$\Phi_{n}^{l(\alpha)}(\phi_{\alpha},x)$. 
In the actual numerical calculations,  
we use $\tilde{U}^{l}_{\alpha}(\phi_{+},x,0)$ and 
$\tilde{U}^{l}_{\alpha}(\phi_{+},0,x)$
given by 
\begin{align}
\tilde{U}^{\rm N}_{+}(\phi_{+},x,0)
& = U^{\rm N}_{+}(\phi_{+},x,0)e^{\kappa^{\rm N}_{+}x},
\\
\tilde{U}^{\rm D}_{+}(\phi_{+},0,x)
& = U^{\rm D}_{+}(\phi_{+},0,x)e^{-\kappa^{\rm D}_{+}x}.
\end{align}
\par
\section{For avoiding divergence
in our calculation}
\label{app:03}
In particular,
if $\Delta^{\rm N}(\phi_{+},-\infty)$
$[\Delta^{\rm D}(\phi_{-}, \infty)]=0$,
we can find ${\cal G}^{\rm N}_{+}$
$[{\cal G}^{\rm D}_{-}] \rightarrow \infty$.
In this case, instead of Eq.~(\ref{req}),
we solve the following equation as
\begin{align}
\hbar |v_{{\rm F}x}|
\frac{\partial}{\partial x}
\left ( \frac{1}
{{\cal G}^{l}_{\alpha}(x)}
\right )
=& -\alpha \left [
2\omega_{m} \left ( \frac{1}
{{\cal G}^{l}_{\alpha}(x)}
\right ) \right . \nonumber \\
+ \Delta^{l}(\phi_{\alpha},x)^{*}
& \left . \left ( \frac{1}
{{\cal G}^{l}_{\alpha}(x)}
\right )^{2}
- \Delta^{l}(\phi_{\alpha},x)
\right ],
\end{align}
under initial condition,
\begin{align}
\frac{1}{{\cal G}^{l}_{\alpha}(x)}
= - \frac{\Delta^{l}(\phi_{\alpha},-\alpha \infty)}
{\omega_{m}+ \Omega^{l}_{\alpha}}.
\end{align}



\begin{thebibliography}{99}
\bibitem{SR92}
M. Sigrist and T.M. Rice,
J. Phys. Soc. Jpn. {\bf 61}, 4283 (1992).
%
\bibitem{SR95}
M. Sigrist and T.M. Rice,
Rev. Mod. Phys. {\bf 67}, 505 (1995).
%
\bibitem{Scalapino}
D.J. Scalapino,
Phys. Rep. {\bf 250}, 329 (1995).
%
\bibitem{Harlin}
D.J. Van Harlingen,
Rev. Mod. Phys. {\bf 67}, 515 (1995).
%
\bibitem{Tsuei}
C.C. Tsuei and J.R. Kirtley, 
Rev. Mod. Phys. {\bf 72}, 969 (2001).




\bibitem{Buch}
L.J. Buchholtz and G. Zwicknagl,
Phys. Rev. B {\bf 23}, 5788 (1981). 
%
\bibitem{Hu}
C.R. Hu,
Phys. Rev. Lett. {\bf 72}, 1526 (1994).
%
\bibitem{TK95}
Y. Tanaka and S. Kashiwaya,
Phys. Rev. Lett. {\bf 74}, 3451 (1995),
%
\bibitem{KT95}
S. Kashiwaya, Y. Tanaka, M. Koyanagi, H. Takashima,
and K. Kajimura, Phys. Rev. B {\bf 51}, 1350 (1995).
%
\bibitem{Matsu1}
M. Matsumoto and H. Shiba,
J. Phys. Soc. Jpn. {\bf 64}, 3384 (1995).
%
\bibitem{FRS97}
M. Fogelstr\"{o}m, D. Rainer, and J.A. Sauls,
Phys. Rev. Lett. {\bf 79}, 281 (1997).
%
\bibitem{KT00}
S. Kashiwaya and Y. Tanaka,
Rep. Prog. Phys. {\bf 63}, 1641 (2000).
%
\bibitem{KT96}
S. Kashiwaya, Y. Tanaka, 
M. Koyanagi, and K. Kajimura,
Phys. Rev. B {\bf 53}, 2667 (1996).
%
\bibitem{Barash}
Yu. S. Barash, A. A. Svidzinsky, and H. Burkhardt,
Phys. Rev. B {\bf 55}, 15282 (1997).
%
\bibitem{Lofwander}
T. L\"{o}fwander, V.S. Shumeiko, and G. Wendin,
Supercond. Sci. Technol. {\bf 14}, R53 (2001).

\bibitem{ATS04}
Y. Asano, Y. Tanaka and S. Kashiwaya,  
Phys. Rev. B {\bf 69}, 134501 (2004).
%


\bibitem{Tanu1}
Y. Tanuma,
Y. Tanaka, M. Yamashiro, and S. Kashiwaya,
Phys. Rev. B {\bf 57}, 7997 (1998).
%
\bibitem{Tanu2}
Y. Tanuma, Y. Tanaka, M. Ogata, and S. Kashiwaya,
J. Phys. Soc. Jpn. {\bf 67}, 1118 (1998).
%
\bibitem{Tanu2'}
Y. Tanuma, Y. Tanaka, M. Ogata, and S. Kashiwaya,
Phys. Rev. B {\bf 60}, 9817 (1999).
%
\bibitem{Zhu}
J. X. Zhu and C. S. Ting,
Phys. Rev. B {\bf 59}, R14165 (1999).
%
\bibitem{W1}
K.V. Samokhin and M.B. Walker,
Phys. Rev. B {\bf 64}, 172506 (2001).
%
\bibitem{W2}
P. Pairor and M.B. Walker
Phys. Rev. B {\bf 65}, 064507 (2002).
%


%
\bibitem{asa1}
Y. Asano and Y. Tanaka,
Phys. Rev. B {\bf 65}, 064522 (2002).
%
\bibitem{asa2}
Y. Asano, Y. Tanaka, and S. Kashiwaya.
Phys. Rev. B {\bf 69}, 214509 (2004).


\bibitem{r1}
Y. Tanaka, Y. Tanuma, and S. Kashiwaya,
Phys. Rev. B {\bf 64}, (2001) 054510.
%
\bibitem{r2}
Y. Tanaka, Yu.V. Nazarov, and S. Kashiwaya,
Phys. Rev. Lett. {\bf 90}, (2003) 167003.
%
\bibitem{r3}
Y. Tanaka, A.A. Golubov, and S. Kashiwaya,
Phys. Rev. B {\bf 68}, 054513 (2003).
%
\bibitem{r4}
N. Kitaura, H. Itoh, Y. Asano, Y. Tanaka, J. Inoue,
Y. Tanuma, and S. Kashiwaya,
J. Phys. Soc. Jpn. {\bf 72}, 1718 (2003).
%
\bibitem{r5}
A.A. Golubov and M.Y. Kupriyanov,
Pis'ma Zh. Eksp. Teor. fiz {\bf 67}, 478 (1998)
[Sov. Phys. JETP Lett. {\bf 67}, 501 (1998)].
%
\bibitem{r6}
A.A. Golubov and M.Y. Kupriyanov,
Pis'ma Zh. Eksp. Teor. fiz {\bf 69}, 242 (1999)
[Sov. Phys. JETP Lett. {\bf 69}, 262 (1999)].
%
\bibitem{r7}
A. Poenicke, Yu.S. Barash, C. Bruder, and V. Istyukov,
Phys. Rev. B {\bf 59}, 7102 (1999).
%
\bibitem{r8}
K. Yamada, Y. Nagato, S. Higashitani, and K. Nagai,
J. Phys. Soc. Jpn. {\bf 65}, 1540 (1996).
%
\bibitem{r9}
T. L\"{u}ck, U. Eckern, and A. Shelankov,
Phys. Rev. B {\bf 63}, 064510 (2002).
%
\bibitem{Lubi03}
I. Lubimova and G. Koren,
Phys. Rev. B \textbf{68}, 224519 (2003).

\bibitem{e1}
J. Geerk, X.X. Xi, and G. Linker,
Z. Phys. B. {\bf 73}, 329 (1988).
%
\bibitem{e2}
J. Lesueur, L.H. Greene, W.L. Feldman,
and A. Inam,
Physica C {\bf 191}, 325 (1992).
%
\bibitem{e3}
S. Kashiwaya, Y. Tanaka, M. Koyanagi,
H. Takashima, and K. Kajimura,
Phys. Rev. B {\bf 51}, 1350 (1995).
%
\bibitem{e4}
L. Alff, H. Takashima, S. Kashiwaya, N. Terada, H. Ihara,
Y. Tanaka, M. Koyanagi, and K. Kajimura,
Phys. Rev. B {\bf 55}, R14757 (1997).
%
\bibitem{e5}
M. Covington, M. Aprili, E. Paraoanu,
L.H. Greene, F. Xu, J. Zhu,
and C.A. Mirkin,
Phys. Rev. Lett. {\bf 79}, 277 (1997).
%
\bibitem{e6}
H. Aubin, L. H. Greene, Sha Jian, and D. G. Hinks,
Phys. Rev. Lett. {\bf 89}, 177001 (2002).
%
\bibitem{e7}
J.Y.T. Wei, N.-C. Yeh, D.F. Garrigus, and M. Strasik,
Phys. Rev. Lett. {\bf 81}, 2542 (1998).
%
\bibitem{e8}
I. Iguchi, W. Wang, M. Yamazaki, Y. Tanaka,
and S. Kashiwaya,
Phys. Rev. B {\bf 62}, R6131 (2000).
%
\bibitem{e9}
Y. Dagan and G. Deutscher,
Phys. Rev. Lett. {\bf 87}, 177004 (2001).
%
\bibitem{e10}
A. Sharoni, O. Millo, A. Kohen, Y. Dagan, R. Beck,
G. Deutscher, and G. Koren,
Phys. Rev. B {\bf 65}, 134526 (2002).
%
\bibitem{Koren}
G. Koren, L. Shkedy, and E. Polturak,
cond-mat/0306594 (2003).
%
\bibitem{Boston}
S. Kashiwaya, Y. Tanaka, N. Terada, M. Koyanagi,
S. Ueno, L. Alff, H. Takashima, 
Y. Tanuma, and K. Kajimura,
J. Phys. Chem. Solids. {\bf 59}, 2034 (1998).
%
\bibitem{e11}
J.W. Ekin, Y. Xu, S. Mao, T. Venkatesan, D.W. Face,
M. Eddy, and S. A. Wolf,
Phys. Rev. B {\bf 56}, 13746 (1997).
%
\bibitem{e12}
A. Biswas, P. Fournier, M.M. Qazilbash,
V.N. Smolyaninova, H. Balci, and R.L. Greene,
Phys. Rev. Lett. {\bf 88}, 207004 (2002).
%
\bibitem{e13}
M.M. Qazilbash, A. Biswas, Y. Dagan,
R.A. Ott, and R.L. Greene,
Phys. Rev. B {\bf 68}, 024502 (2003).
%

\bibitem{SBL95}
M. Sigrist, D. B. Bailey, and R. B. Laughlin,
Phys. Rev. Lett. {\bf 74}, 3249 (1995).
%
\bibitem{Kuboki}
K. Kuboki and M. Sigrist,
J. Phys. Soc. Jpn. {\bf 65}, 361 (1996).
%
\bibitem{Sig98}
M. Sigrist,
Prog. Theo. Phys. {\bf 99}, 899 (1998).
%
\bibitem{Matsu2}
M. Matsumoto and H. Shiba,
J. Phys. Soc. Jpn. {\bf 64}, 4867 (1995).
%
\bibitem{Matsu3}
M. Matsumoto and H. Shiba,
J. Phys. Soc. Jpn. {\bf 65}, 2194 (1996).
%
\bibitem{Krishana}
K. Krishana, N. P. Ong, Q. Li,
G. D. Gu, and N. Koshizuka,
Science {\bf 277}, 83 (1997).
%
\bibitem{Balatsky}
A. V. Balatsky,
Phys. Rev. Lett. {\bf 80}, 1972 (1998).
%
\bibitem{Laughlin}
R. B. Laughlin,
Phys. Rev. Lett. {\bf 80}, 5188 (1998).

%
\bibitem{Neils}
W.K. Neils and D.J. Van Harlingen,
Phys. Rev. Lett. {\bf 88}, 047001 (2002).
%
\bibitem{Tanu5}
Y. Tanuma, Y. Tanaka, and S. Kashiwaya,
Phys. Rev. B {\bf 64}, 214519 (2001).
%

\bibitem{Kohen}
A. Kohen, G. Leibovitch, and G. Deutscher,
Phys. Rev. Lett. {\bf 90}, 207005 (2003).
%
%
\bibitem{Ohashi}
Y. Ohashi,
J. Phys. Soc. {\bf 65}, 823 (1996).

\bibitem{Lof}
T. L\"{o}fwander,
Phys. Rev. B {\bf 70}, 094518 (2004).




\bibitem{Eilen}
G. Eilenberger,
Z. Phys. \textbf{214}, 195 (1968).
%
\bibitem{Nagai}
K. Nagai,
\textit{Quasiclassical Methods
in Superconductivity and Superfluidity},
edited by D. Rainer and J.A. Sauls
(unpublished).
%
\bibitem{Ashida}
M. Ashida, S. Aoyama, J. Hara, and K. Nagai,
Phys. Rev. B \textbf{40}, 8673 (1989).
%
\bibitem{Nagato}
Y. Nagato, K. Nagai, and J. Hara,
J. Low Temp. Phys. {\bf 93}, 33 (1993).
%
\bibitem{Bruder}
C. Bruder,
Phys. Rev. B \textbf{41}, 4017 (1990).
%
\bibitem{Kurki}
J. Kurkij\"{a}rvi and D. Rainer,
\textit{Helium Three},
edited by W. P. Halperin and L. P. Pitaevskii
(Elsevier, Amsterdam, 1990).

\bibitem{Tanaka9371}
Y. Tanaka and S. Kashiwaya, Phys. Rev. B {\bf 53} 9371 (1996). 
%
\bibitem{Tanaka2004}
Y. Tanaka, Y.V. Nazarov, A.A. Golubov, and S. Kashiwaya,
Phys. Rev. B  {\bf 69} 144519 (2004), 
Y. Tanaka and S. Kashiwaya,
Phys. Rev. B {\bf 70} 012507 (2004).
%
\bibitem{a}
M. Yamashiro, Y. Tanaka, and S. Kashiwaya,
Phys. Rev. B {\bf 56}, 7847 (1997).
%
\bibitem{b}
M. Yamashiro, Y. Tanaka, and S. Kashiwaya,
J. Phys. Soc. Jpn. {\bf 67}, 3364 (1998).
%
\bibitem{c}
M. Yamashiro, Y. Tanaka Y. Tanuma, and S. Kashiwaya,
J. Phys. Soc. Jpn. {\bf 67}, 3224 (1998).
%
\bibitem{d}
M. Yamashiro, Y. Tanaka Y. Tanuma, and S. Kashiwaya,
J. Phys. Soc. Jpn. {\bf 68}, 2019 (1999).
%
\bibitem{e}
C. Honerkamp and M. Sigrist,
Prog. Theor. Phys. {\bf 100}, 53 (1998).
%
\bibitem{f}
C. Honerkamp and M. Sigrist,
J. Low Temp. Phys. {\bf 111}, 895 (1998).
%
\bibitem{g}
Y. Tanaka, T. Asai, N. Yoshida, J. Inoue, and S. Kashiwaya,
Phys. Rev. B  {\bf 61}, R11902 (2000).
%
\bibitem{h}
Y. Tanaka, T. Hirai, K. Kusakabe, and S. Kashiwaya,
Phys. Rev. B {\bf 60}, 6308 (1999).
%
\bibitem{i}
T. Hirai, K. Kusakabe, and Y. Tanaka,
Physica C {\bf 336}, 107 (2000).
%
\bibitem{j}
K. Kusakabe and Y. Tanaka,
Physica C {\bf 367}, 123 (2002).
%
\bibitem{k}
K. Kusakabe and Y. Tanaka,
J. Phys. Chem. Solids {\bf 63}, 1511 (2002).
%
\bibitem{l}
N. Stefanakis,
Phys. Rev. B {\bf 64}, 224502 (2001).
%
\bibitem{m}
Z.C. Dong, D.Y. Xing, and Jinming Dong,
Phys. Rev. B {\bf 65}, 214512 (2002).
%
\bibitem{n}
Z.C. Dong, D.Y. Xing, Z.D. Wang,
Ziming Zheng, and Jinming Dong,
Phys. Rev. B {\bf 63}, 144520 (2001).
%
\bibitem{o}
Yu.S. Barash, M.S. Kalenkov, and J. Kurkij\"{a}rvi,
Phys. Rev. B {\bf 62}, 6665 (2000).
%
\bibitem{p}
M.H.S. Amin, A.N. Omelyanchouk, and A.M. Zagoskin,
Phys. Rev. B {\bf 63}, 212502 (2001).
%
\bibitem{q}
Shin-Tza Wu and Chung-Yu Mou,
Phys. Rev. B {\bf 66}, 012512 (2002).
%
\bibitem{r}
J.-X. Zhu, B. Friedman, and C.S. Ting,
Phys. Rev. B {\bf 59}, 9558 (1999).
%
\bibitem{s}
S. Kashiwaya, Y. Tanaka, N. Yoshida, and M.R. Beasley,
Phys. Rev. B {\bf 60}, 3572 (1999).
%
\bibitem{t}
N. Yoshida, Y. Tanaka, J. Inoue, and S. Kashiwaya,
J. Phys. Soc. Jpn. {\bf 68}, 1071 (1999).
%
\bibitem{u}
I. \v{Z}uti\'c and O.T. Valls,
Phys. Rev. B {\bf 60}, 6320 (1999).
%
\bibitem{v}
T. Hirai, N. Yoshida, Y. Tanaka, J. Inoue, and S. Kashiwaya,
J. Phys. Soc. Jpn. {\bf 70}, 1885 (2001).
%
\bibitem{w}
N. Yoshida, H. Itoh, T. Hirai, Y. Tanaka,
J. Inoue, and S. Kashiwaya,
Phsica C {\bf 367}, 135 (2002).
%
\bibitem{x}
T. Hirai, Y. Tanaka, N. Yoshida, Y. Asano, 
J. Inoue, and S. Kashiwaya,
Phys. Rev. B {\bf 67}, 174501 (2003).






\bibitem{o1}
K. Sengupta, I. \v{Z}uti\'c, H.-J. Kwon, V.M. Yakovenko,
and S. Das Sarma,
Phys. Rev. B {\bf 63}, 144531 (2001).
%
\bibitem{o2}
Y. Tanuma, K. Kuroki, Y. Tanaka, and S. Kashiwaya,
Phys. Rev. B {\bf 64}, 214510 (2001).
%
\bibitem{o3}
Y. Tanuma, K. Kuroki, Y. Tanaka, R. Arita,
S. Kashiwaya, and H. Aoki,
Phys. Rev. B {\bf 66}, 094507 (2002).
%
\bibitem{o4}
Y. Tanuma, Y. Tanaka, K. Kuroki,
and S. Kashiwaya,
Phys. Rev. B {\bf 66}, 174502 (2002).
%
\bibitem{o5}
Y. Tanuma, K. Kuroki, Y. Tanaka,
and S. Kashiwaya,
Phys. Rev. B {\bf 68}, 214513 (2003).




%
\bibitem{K95}
S. Kashiwaya, Y. Tanaka, M. Koyanagi,
and K. Kajimura,
J. Phys. Chem. Solids {\bf 56}, 1721 (1995). 
%


%
\bibitem{m1}
Y. Tanaka, H. Tsuchiura, Y. Tanuma, and S. Kashiwaya,
J. Phys. Soc. Jpn. {\bf 71}, 271 (2002).
%
\bibitem{m2}
Y. Tanaka, Y. Tanuma  K. Kuroki, and S. Kashiwaya,
J. Phys. Soc. Jpn. {\bf 71}, 2102 (2002).
%
\bibitem{m3}
Y. Tanaka, H. Itoh, H. Tsuchiura, Y. Tanuma,
J. Inoue, and S. Kashiwya,
J. Phys. Soc.  Jpn. {\bf 71}, 2005 (2002).
%

%
\bibitem{J1}
Y. Tanaka and S. Kashiwaya,
Phys. Rev. B {\bf 53}, R11957 (1996).
%
\bibitem{J2}
Y. Tanaka and S. Kashiwaya,
Phys. Rev. B {\bf 56}, 892 (1997).
%
\bibitem{J3}
Y. Tanaka and S. Kashiwaya,
Phys. Rev. B {\bf 58}, R2948 (1998).
%
\bibitem{J4}
Y. Tanaka and S. Kashiwaya,
J. Phys. Soc. Jpn. {\bf 68}, 3485 (1999).
%
\bibitem{J5}
Y. Tanaka and S. Kashiwaya,
J. Phys. Soc. Jpn. {\bf 69}, 1152 (2000).

%
\bibitem{J6}
Yu.S. Barash, H. Burkhardt, and D. Rainer,
Phys. Rev. Lett. {\bf 77}, 4070 (1996).
%

\bibitem{J7}
H. Hilgenkamp, J. Mannhart, and B. Mayer,
Phys. Rev. B {\bf 53}, 14586 (1996).


\bibitem{J8}
H. Hilgenkamp and J. Mannhart,
Rev. Mod. Phys. {\bf 74}, 485 (2002).

\bibitem{J9}
E. Il'ichev, M. Grajcar, R. Hlubina, R.P.J. Ijsselsteijn,
H.E. Hoenig, H.-G. Meyer, A. Golubov,
M.H.S. Amin, A.M. Zagoskin,
A.N. Omelyanchouk, and M.Y. Kupriyanov, 
Phys. Rev. Lett. {\bf 86}, 5369 (2001).
%
\bibitem{J10}
E. Il'ichev, V. Zakosarenko, R.P.J. Ijsselsteijn,
V. Schultze, H.-G. Meyer, H.E. Hoenig,
H. Hilgenkamp, and J. Mannhart, 
Phys. Rev. Lett. {\bf 81}, 894 (1998).
%


%
\bibitem{J11}
F. Lombardi, F. Tafuri, F. Ricci,
F. Miletto Granozio, A. Barone, G. Testa,
E. Sarnelli, J.R. Kirtley, and C.C. Tsuei,
Phys. Rev. Lett. {\bf 89}, 207001 (2002).
%
\bibitem{J12}
H.J.H. Smilde, Ariando, D.H.A. Blank,
G.J. Gerritsma, H. Hilgenkamp, and H. Rogalla,
Phys. Rev. Lett. {\bf 88}, 057004 (2002).
%
\bibitem{J13}
T. Imaizumi, T. Kawai, T. Uchiyama, and I. Iguchi,
Phys. Rev. Lett. {\bf 89}, 017005 (2002).
%
\bibitem{J14}
H. Arie, K. Yasuda, H. Kobayashi, I. Iguchi,
Y. Tanaka, and S. Kashiwaya,
Phys. Rev. B {\bf 62}, 11864 (2000).

%
\bibitem{J15}
Y. Asano,
Phys. Rev. B {\bf 63}, 052512 (2001).
%
\bibitem{J16}
Y. Asano,
Phys. Rev. B {\bf 64}, 014511 (2001).
%
\bibitem{J17}
Y. Asano,
Phys. Rev. B {\bf 64}, 224515 (2001).
%
\bibitem{J18}
Y. Asano,
J. Phys. Soc. Jpn. {\bf 71}, 905 (2002).
%
\bibitem{J19}
Y. Asano, Y. Tanaka, M. Sigrist, and S. Kashiwaya,
Phys. Rev. B  {\bf 67}, 184505 (2003). 



\end{thebibliography}
\end{document}